\documentclass[aps,amsmath,amssymb,amsfonts,showpacs,showkeys]{revtex4}

\usepackage{epsfig,psfrag}

\newcommand{\sla}[1]{\mbox{$#1\!\!\!/$}}
\newcommand{\beq}{\begin{equation}}
\newcommand{\eeq}{\end{equation}}
\newcommand{\bea}{\begin{eqnarray}}
\newcommand{\eea}{\end{eqnarray}}
\newcommand{\hf} {\frac{1}{2}}
\newcommand{\nn}{\nonumber\\}

\newcommand\eqn[1]{Eq.\,(\ref{#1})}

\newcommand\fig[1]{Fig.\,{\ref{#1}}}
\newcommand\sect[1]{Sect.\,{\ref{#1}}}

\newcommand\tab[1]{Table~\ref{#1}}
\def\t{\tilde}
\def\tV{\tilde V}
\def\ord#1{{\cal O}(#1)}

\begin{document}

\title{Lectures on renormalization and asymptotic safety}

\author{Sandor Nagy}
\affiliation{Department of Theoretical Physics, University of Debrecen,
P.O. Box 5, H-4010 Debrecen, Hungary}
\affiliation{MTA-DE Particle Physics Research Group, P.O.Box 51,
H-4001 Debrecen, Hungary}
\email{nagys@phys.unideb.hu}

\begin{abstract}

A short introduction is given on the functional renormalization group method,
putting emphasis on its nonperturbative aspects. The method enables to find
nontrivial fixed points in quantum field theoretic models which make them
free from divergences and leads to the concept of asymptotic safety. 
It can be considered as a generalization of the asymptotic
freedom which plays a key role in the perturbative renormalization.
We summarize and give a short discussion of some important models, which are
asymptotically safe such as the Gross-Neveu model, the nonlinear $\sigma$ model,
the sine-Gordon model, and we consider the model of quantum Einstein gravity
which seems to show asymptotic safety, too.
We also give a detailed analysis of infrared behavior of such scalar models
where a spontaneous symmetry breaking takes place. The deep infrared behavior
of the broken phase cannot be treated within the framework of perturbative calculations.
We demonstrate that there exists an infrared fixed point in the broken phase which
creates a new scaling regime there, however its structure is hidden by the singularity
of the renormalization group equations. The theory spaces of these models show
several similar properties, namely the models have the same phase and fixed point
structure. The quantum Einstein gravity also exhibits similarities when considering the
global aspects of its theory space since the appearing two phases there show analogies with
the symmetric and the broken phases of the scalar models.
These results be nicely uncovered by the functional renormalization group method.
\end{abstract}
\maketitle

\section{Introduction}

The concept of asymptotic safety can be understood in the framework of
quantum field theory, where we treat quantum systems with
many degrees of freedom and many orders of magnitude in length, or in energy, momentum.
Usually we have an assumption of describing the high energy/ultra violet (UV),
or short distance/microscopic interactions among the elementary particles where
a high degree of symmetry exists. However the measurements are typically
performed at low energies, in the infrared (IR) regime, therefore we need the theory,
where those quantum fluctuations or modes are taken into account which energy is between
the UV and IR energy scales. The functional renormalization group (RG) method is one of the
best candidates to take into account the quantum fluctuations step by step, one by one,
systematically \cite{Wetterich:1992yh,Berges:2000ew,Polonyi:2001se,Pawlowski:2005xe,Gies:2006wv,
Delamotte:2007pf,Rosten:2010vm}.
The method provides us a partial differential equation for the
action of the model. As an initial condition we need the UV action, and the
solution of the equation gives us the IR one which can be identified with the effective
potential. Generally the IR behavior of a model can differ significantly from the UV one.

Starting from the classical UV (or blocked) action one can create the generating functional of the
connected correlation functions which Legendre transform gives the effective action. We assume
(especially in scalar models), that the blocked and the effective action has the same
functional structure. We usually perform a field expansion to the effective action
which results in many interacting terms, and each of them is multiplied by a coupling. The
original partial differential equation can be converted to a system of ordinary
differential equations w.r.t. the couplings. These equations are called evolution or
flow equations. The initial conditions are the UV values of the couplings and the
results of the equations are the IR ones. The RG method typically gives highly
nonlinear evolution equations, giving flows which cannot be recovered by a perturbative
RG treatment.

Moreover the perturbative RG calculations are usually performed in the vicinity of
the Gaussian fixed point (GFP). The GFP is the origin of the theory space which is spanned
by the (dimensionless) couplings. The UV effective action is given by small values of
the initial couplings, so the kinetic term dominates the UV physics. If the UV scaling
of the couplings are UV attractive or IR repulsive (they go away from the fixed point if we
consider the direction of the RG 'time' scale) then the perturbative RG equations of the model
keeps the finiteness of the couplings. Otherwise it may happen that the GFP is a hyperbolic point which means
that there are directions in the theory space which are UV attracted by the GFP but there
are certain ones which are UV repulsed. In the latter case the perturbative flow
equations give that the corresponding couplings go to infinity, which seems nonphysical. However
the model can be safe from divergences if besides the GFP there exists a further nontrivial
fixed point in the finite region of the theory space which UV attracts the trajectories, as
the GFP does usually in the perturbatively renormalizable case. It can make the physical quantities
finite, due to the finiteness of the dimensionless couplings. The nontrivial fixed point
is usually referred to as a non Gaussian fixed point (NGFP). This is the main idea of
asymptotic safety, Shortly, it means that there exists a UV attractive NGFP in the theory space.

The paper is organized as follows.
In \sect{sect:ren}  we give a short introduction on the functional RG method with some
basic concepts and tools. Then the d-dimensional $O(N)$ model is investigated as a classical
example of the asymptotic freedom. In \sect{sect:as} we treat the asymptotic safety, as the
generalization of the idea of the asymptotic freedom. It is shown in the framework of the
nonlinear $\sigma$, Gross-Neveu, sine Gordon and quantum Einstein gravity models how the
asymptotic safety takes place. In \sect{sect:sum} the conclusions are drawn up.

\section{Renormalization}\label{sect:ren}

The functional renormalization group method is one of the most useful nonperturbative
tools to investigate quantum field theoretic models.
We define a model by its action at a high (UV) energy scale, because at UV the form
of the action is generally quite simple due to the high symmetries appearing at large energies,
which belongs to the short distance microscopic interactions. The UV action contains
interaction terms multiplied by the UV couplings. The RG method systematically eliminates
the degrees of freedom of the theory in order of decreasing energy, and provides us the value
of the couplings at lower energy scales. As a result we obtain the value of IR couplings
at practically zero energies.
In the IR the effective potential usually has an involved structure. There the long-range
interactions may induce further couplings and can induce non-localities, or global
condensates.

The nonperturbative nature of the functional RG can be nicely enlightened, if one
considers the tunneling, which is a highly nonperturbative phenomenon in Quantum Mechanics.
In the path integral formulation of QFT the extremum of the UV action is the classical path.
Thus if one treats QM then the tree level RG evolution corresponds to this
classical path. It implies that the RG method
should dress up the couplings in such a way which can drive the classical path to a path
which corresponds to the tunneling process. Naturally it requires all the modes till the deep IR
regime \cite{Horikoshi:1998sw,Kapoyannis:2000sp,Zappala:2001nv,2002PThPh.108..571A,Nagy:2010fv}.
Moreover we need further terms in the gradient expansion. We note that the
tunneling can be treated by a more exact manner in the framework of solving Schr\"odinger
equation with the corresponding potential numerically. However the Schr\"odinger equation
can be considered as an effective theory, while the path integral formulation of QFT
is more fundamental \cite{Polonyi:1994pn}. The wavefunction which corresponds to the tunneling should be combined
from practically plane waves since the field variables in QFT are expanded from them.
If one includes higher order terms in the derivative expansion then more involved
functions can appear in the field expansion. This fact requires to take into account
all the modes and the wavefunction renormalization. Both can be systematically treated
in the framework of the RG method.

The quantized anharmonic oscillator coupled to a heat bath provides us a system, where
the ``holy grail'' of the phase transitions, namely the quantum-classical transition
can be investigated \cite{Aoki:1999rt,Aoki:2002yt,Aoki:2002ax,Nagy:2006apd,Polonyi:2011yx,
2012arXiv1209.5827A}.
The heat bath can be considered as the environment of the original quantized oscillator,
which can be identified by the system. The environment can be represented by harmonic
oscillators which are coupled to the system. The RG procedure integrates out
the environment and leads to an effective system. It appears as if the quantum effects
are dissipated by the environment driving the system to a classical regime.

The RG method can be used in a powerful manner in almost every area of modern physics.
In condensed matter systems, e.g. one can investigate the Bose-Einstein condensate -- BCS
superconductor transition for ultracold fermionic atoms \cite{Diehl:2005an,Diehl:2007th,
Metzner:2011cw}.  Furthermore the RG method can account for the essential scaling of the
correlation length, which typically appears in low dimensional thin superfluid
film structures. The appearing Kosterlitz-Thouless (KT) type or infinite order phase
transitions \cite{Berezinskii:1971,Kosterlitz:1973xp} can be described in the framework of the
2d $O(2)$ model \cite{Grater:1994qx,
VonGersdorff:2000kp}, in the 2d sine-Gordon (SG) model, too \cite{Amit:1979ab,Balog:2000qr,
Nagy:2009pj,Nagy:2010mf}, or its generalization in fermionic models
\cite{Kaplan:2009kr,Braun:2010qs}.

New ideas or improvements of the RG method are usually investigated in the d-dimensional
$O(N)$ model. The calculation of the critical exponents in the 3d $O(N)$ model provides us a
widely acceptable testing ground of the functional renormalization. 
One can get the exponents by field expansion of the potential \cite{Tetradis:1993ts,Liao:1999sh,
Litim:2001up}, or by $\epsilon$-expansion in $4-\epsilon$ dimensions
\cite{Guida:1996ep,Guida:1998bx,ZinnJustin:1999bf}. The convergence
of the exponents in the derivative expansion is also investigated
\cite{Morris:1996xq,Morris:1997xj,Litim:2010tt,Canet:2002gs,Canet:2003qd}, however
preciser results can be obtained without expanding the potential
\cite{Bervillier:2005za,Pangon:2009pj,Pangon2009}.
Furthermore, the supersymmetric version of the model has also been attracted a considerable
attention \cite{Litim:2011bf,Heilmann:2012yf}.
The exponents can be determined from IR scalings, too \cite{Nagy:2012np}. Likewise,
less computational effort can give very good results for the exponents by using
the BMW approximation, where the full momentum dependence of the correlation functions
is considered \cite{Benitez:2007mk,Benitez:2009xg,Benitez:2011xx}.

The RG treatment of 2+1 fermionic systems has a revival since the new results on graphene.
The Gross-Neveu model \cite{Castorina:2003kq,Braun:2011pp,Scherer:2012nn}
give on the other hand a good testing
ground to investigate a model with NGFP and asymptotic safety. The Thirring model
\cite{Gies:2010st,Janssen:2012pq} serves as a good playing ground to investigate the
chiral symmetry breaking, which appears in electroweak interactions.
The nonperturbative treatment of the RG method is capable of describing quantum
electrodynamics \cite{Alexandre:2001wj,Gies:2004hy,Arnone:2005vd},
non-abelian gauge theories \cite{Litim:1998qi,D'Attanasio:1996jd,Morris:1999px,
Morris:2000fs,Arnone:2002cs,Arnone:2005fb,Morris:2005tv,Patkos:2012ex}, furthermore the
confining mechanism in quantum chromodynamics even in finite temperature and chemical potential
\cite{Litim:2002ce,Pawlowski:2003hq,Morris:2006in,Braun:2009gm}. 

The nonlinear $\sigma$ model itself has a wide broad interest in many branches in
quantum physics starting from phenomenological aspects of high energy physics,
condensed matter systems, and strings. Furthermore it exhibits a symmetric and a broken
symmetric phase, similarly to the $O(N)$ model and the quantum Einstein gravity (QEG),
and the model also has a NGFP in the UV \cite{Codello:2008qq,Percacci:2009fh,Percacci:2009dt,
Fabbrichesi:2010xy,Percacci:2012mx}. The other common feature with QEG is that they
have nonpolynomial interactions, the couplings have the same dimension there, and both
models need background field technique to construct the RG equations \cite{Percacci:2009dt}
which is widely used in gauge theories \cite{Litim:2002hj} and in QEG \cite{Reuter:1996cp}.
Among the low dimensional scalar models we also mention the 2d SG model
\cite{Coleman:1974bu}, which is non-trivial quantum field theory with compact
variables -- similarly to the non-Abelian gauge theories -- which is supposed to be the key to
the confinement mechanism. The functional RG treatment shows that the SG model has two phases
\cite{Nandori:1999vi,Nagy:2006pq,Nagy:2009pj,Nagy:2010mf} and there is an IR fixed point
in the broken phase \cite{Nagy:2009pj,Nagy:2010mf}, furthermore the model has both a UV
Gaussian and a NGFP, and what is more, the latter shows singularity.

The IR limit of the RG flows is also a great challenge to reach. In scalar models
the IR scaling of the symmetric phase can be easily obtained. It can be
characterized by such an effective potential that has a well defined single minimum at
the origin at the value of the field variable $\phi=0$.
In the phase with spontaneously broken symmetry the effective
potential tends to be degenerate which means that a plateau starts to form in the
potential around the origin \cite{Tetradis:1992qt,Berges:2000ew,Delamotte:2007pf}.
It is due to the huge amount of soft (nearly zero energy)
modes which can excite the ground state without energy practically
\cite{Alexandre:1997gj,Alexandre:1998ts,Polonyi:2001se}. In this region
of the theory space the perturbative treatments naturally do not work. However the RG method
uncovers us that there is an IR fixed point in the broken phase
\cite{Tetradis:1992qt,Nagy:2009pj,Kaplan:2009kr,Nagy:2010mf,Braun:2010tt}. One can define
the correlation length in the IR regime which can help us to determine the corresponding
critical exponent $\nu$, moreover one can also determine the order of the phase transition
of the model \cite{Nagy:2012np,Nagy:2012qz,Nagy:2012rn}.
The IR fixed point explicitly shows the limitation of the theory where it can be treated by
its original degrees of freedom. The bulk amount of soft modes show that
new elementary excitations arises in the model at low energies and its treatment needs a
new model, or at least some new interaction terms.

The other limit towards the UV gives a further challenge of the RG method.
Recently the UV limit of QEG is widely investigated
\cite{Reuter:1996cp,Lauscher:2001rz,Reuter:2001ag,Ashtekar:2004eh,
Reuter:2007rv,kiefer2007quantum,Reuter:2012id}.
The model usually has a GFP. However around the GFP the Newton constant or coupling starts to
blow up in the UV limit, which makes the physical quantities infinitely large there.
It implies that that the model is perturbatively nonrenormalizable
\cite{'tHooft:1974bx,Goroff:1985sz}.
As a possible solution for this problem it was conjectured \cite{Weinberg:1979,Weinberg:1996kw,
Weinberg:2009ca,Weinberg:2009bg} and later showed in
low dimensions \cite{Christensen:1978sc,Gastmans:1977ad},
that there is a further UV NGFP fixed point in the theory space which
makes the Newton constant finite and the model becomes safe from divergences,
which is called asymptotic safety
\cite{Percacci:2003jz,Reuter:2007rv,Percacci:2007sz,Niedermaier:2006ns,Niedermaier:2006wt}.
Sometimes QEG is also referred to as asymptotically safe quantum gravity.
Interestingly the model also has an IR fixed point in its broken phase
\cite{Bonanno:2001hi,Reuter:2002kd,Donkin:2012ud,Nagy:2012rn,Litim:2012vz,Christiansen:2012rx}
at least in the framework of the Einstein-Hilbert truncation.

\subsection{Blocking}

QFT can be formulated by using the path integral formalism \cite{kleinert2004b}.
The generating functional expresses the vacuum--vacuum transition amplitude and has the form
\beq
Z = \int{\cal D}\phi e^{-S_k}=\int d\phi_0\ldots d\phi_{k-\Delta k}d\phi_k d\phi_{k+\Delta k}
\ldots d\phi_\infty e^{-S_k}
\eeq
where $S_k$ is the (blocked) action at the momentum scale $k$. The extremum of the action
gives the classical path. 
The path integral is performed for all the possible field configurations between the given initial and final
states. In order to handle them we order the field configurations according to e.g. by their momentum
\cite{Wegner:1972ih} or even by their amplitude in the internal space \cite{Alexandre:2000eg}.
The procedure leads to the Wetterich equation \cite{Wetterich:1992yh} for the effective
average action \cite{Wetterich:1989xg}. We can assign a certain $k$ to
each field configuration, usually we enumerate them by their decreasing value. We note
that the momentum $k$ is basically a bookkeeping device for the modes, and it has no direct
connection to any value of energy or momentum of the modes. The RG method provides us
a systematic treatment to take into account the quantum fluctuation systematically.
The modes are integrated out one by one in with decreasing value of $k$. After the elimination
of some modes, the action changes to $S_{k-\Delta k}$, which is given at the lower
scale $k-\Delta k$. If one eliminates the modes
in the vicinity of the UV scale, then we recover the perturbative RG treatment.

\begin{center}
\begin{figure}[ht]
\psfrag{ir}{{(IR)~~ $0\leftarrow k$}}
\psfrag{uv}{{$k\to \infty$ ~~(UV)}}
\psfrag{gm}{{$g_{k{-}\Delta k}$}}
\psfrag{gk}{{$g_k$}}
\psfrag{gp}{{$g_{k{+}\Delta k}$}}
\psfrag{km}{{\tiny $k{-}\Delta k$}}
\psfrag{kk}{{\tiny $k$}}
\psfrag{kp}{{\tiny $k{+}\Delta k$}}
\includegraphics[width=10cm]{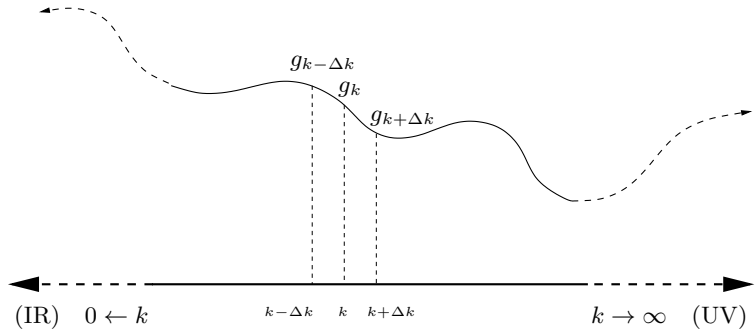}
\caption{\label{fig:scale}
The change of the coupling $g$ is presented as the scale changes from
$k+\Delta k~\to~k~\to~k-\Delta k$. The limit $k\to\infty$ ($k\to 0$) corresponds
to the UV (IR) limits, respectively.}
\end{figure}
\end{center}

The highest scale $k$ is the UV cutoff, which is denoted by $k_\Lambda$. The blocked action
is built up on e.g. symmetry considerations or analogies to other field theoretic models.
Usually the action contains a kinetic and a potential terms, the latter contains interacting
terms that are multiplied by couplings, and they carry the scale dependence in the action.
The couplings usually have physical meaning, they can be related to e.g. particle masses or
interaction couplings. The initial value of the couplings should be given at a high energy
UV scale describing short-range interactions. If one integrates out the modes between the scales
$k$ and $k-\Delta k$ then the value of the coupling changes. It is illustrated in
\fig{fig:scale}, where the coupling is denoted by $g_k$.
The figure illustrates how the value of the coupling changes as the scale $k$ decreases.
The scale $k$ starts at $k_\Lambda$ and tends to zero. One has to integrate
out infinitely many degrees of freedom, where the modes are enumerated by the
continuous index $k$. The coupling becomes a scale dependent function,
the RG evolution eventually gives how the value of the coupling changes
till we reach the scale $k=0$, and then we obtain the effective potential
which contains the coupling $g_0$. From this point of view the functional RG method
can be considered as a tool to perform the path integral in the generating functional.

The RG method gives a partial differential equation for the action, however if one takes a
functional ansatz for its functional form, e.g.
\beq
S_k[\phi,g_i]=\sum_i g_i(k){\cal F}_i(\phi),
\eeq
then one can deduce a system of differential equations for the couplings. Here $k$ denotes
the scale, $g_i(=g_i(k))$ are the dimensionful couplings, and $\phi$ is the field variable.
The functionals ${\cal F}$ are typically Taylor expanded according to its field
variable but there are situations where a Fourier expansion is required if the model
is periodic in the field variable and one would like keep the periodic symmetry.

\subsection{The Wetterich equation}

The RG evolution equation for the blocked action is derived by Wilson \cite{Wilson:1974mb}
and Polchinski \cite{Polchinski:1983gv}. Wilson's idea considers the quantum fluctuation by
decreasing order in the scale $k$, which can correspond to the summation of the
loop expansion. There the short distance interactions are integrated out in order to get an
effective long distance theory. Polchinski's RG method seems to sum up the perturbative
expansion systematically \cite{Polonyi:2001se}.

The Wetterich equation is a functional integro-differential equation for the effective action
\cite{Wetterich:1992yh,Berges:2000ew}. We start its derivation from the generating functional
\beq\label{Zreg}
Z=e^{W_k[J]}=\int {\cal D}[\phi]e^{-(S_k+{\cal R}_k[\phi]-J\cdot \phi)},
\eeq
in Euclidean spacetime, where $J$ denotes the source of the field variable.
We use the notation $f \cdot g=\int_{-\infty}^\infty d^d x f(x)g(x)$.
The integral gives the average of fields over a volume $k^{-1}$. One integrates out
the modes with the scale larger than $k$. There is an additional
term besides the action, which is called the IR regulator, and it has the form
\beq
{\cal R}_k[\phi]=\hf \phi \cdot {\cal R}_k \cdot \phi,
\eeq
which acts as an IR cutoff. Here the dots also denote momentum integrals.
The IR regulator guarantees that the modes with its scale larger than $k$ are taken into
account in an unaltered way while the modes with scale smaller than $k$ are suppressed.
It should satisfy the conditions
\begin{enumerate}
\item $\underset{p^2/k^2\to 0}\lim {\cal R}_k>0$, i.e. it serves as an IR regulator, since
it removes the IR divergences,
\item $\underset{k^2/p^2\to 0}\lim {\cal R}_k\to 0$, which expresses that if the regulator
is removed, then we get back $Z$ when $k\to 0$, so we obtain back the original model
in this limit,
\item $\underset{k^2\to \infty}\lim {\cal R}_k\to \infty$, which ensures that the
microscopic action can be recovered in the limit $S=\lim_{k\to k_\Lambda} \Gamma_k$.
It also serves as a UV regulator.
\end{enumerate}
We have many possibilities to choose a functional from to the IR regulator
\cite{Litim:2000ci,Litim:2001up,Lauscher:2002sq,Pawlowski:2005xe,Lauscher:2001ya,Nandori:2012tc}.
Some typical regulators which are used frequently in the literature are
\bea
\label{regexp}
{\cal R}_k &=& ap^2\frac{e^{-b(p^2/k^2)^c}}{1-e^{-b(p^2/k^2)^c}} ~~\mbox{(exponential),}\\
\label{regpow}{\cal R}_k &=& p^2\left(\frac{k^2}{p^2}\right)^b~~\mbox{(power~law),}\\
\label{regopt}{\cal R}_k &=& (k^2-p^2)\theta(k^2-p^2)~~\mbox{(optimized~or~Litim's).}
\eea
The effective potential should not depend on the IR regulator. However RG equations
cannot be solved without approximations. The introduction of some truncations in the functional ansatz
for the action induces some regulator dependence, therefore one has to check how the obtained flows depend on
them. Differentiating both sides of \eqn{Zreg} by $k$ one obtains
\bea
\partial_k W_k[J] &=& e^{-W_k[J]}
\int {\cal D}[\phi] \partial_k{\cal R}_k e^{-(S_k+{\cal R}_k[\phi]-J\cdot \phi)}\nn
&=&-e^{-W_k[J]}\partial_k {\cal R}_k\left [\frac{\delta }{\delta J}\right]e^{W_k[J]}.
\eea
The effective action $\Gamma_k[\phi]$ is the Legendre transform of $W_k[J]$ according to
\beq
\Gamma_k[\phi]=\sup_J\left(- W_k[J]+J \cdot \phi\right),
\eeq
with the field variable
\beq
\phi=\frac{\delta W[J]}{\delta J}.
\eeq
The derivative of the effective action $\Gamma_k[\phi]$ w.r.t the scale $k$ is
\beq
\partial_k \Gamma_k[\phi]=-\partial_k W_k[J]-\frac{\delta W[J]}{\delta J}\partial_k J
+\partial_k J\phi =-\partial_k W_k[J].
\eeq
We rewrite the effective action according to $\Gamma_k[\phi]+ {\cal R}_k[\phi]\to \Gamma_k$,
introduce the 'RG time' $t=\log\frac{k}{k_0}$ and then we obtain the Wetterich equation
\beq\label{wettRG}
\dot\Gamma_k=\hf\mbox{Tr}\frac{\dot {\cal R}_k}{{\cal R}_k+\Gamma_k''},
\eeq
where the notations  $^\prime=\partial/\partial\varphi$ and
$\dot ~ =\partial/\partial t$ are used and the trace Tr
stands for the integration over all momenta and the summation over the internal indices.
The functional form of the effective action is assumed to be similar to the
microscopic action
\bea
\Gamma_k &\sim& S_k\nn
&=& \sum_i g_i(k){\cal F}_i(\phi).
\eea
Typically the effective action for scalar fields is approximated by the operator expansion according to
\beq\label{effac}
\Gamma_k =\int d^d x \left[\hf Z_k(\phi_x)(\partial_\mu \phi_x)^2+V_k(\phi_x) \right],
\eeq
where we the effective action is constructed by operator terms with increasing mass dimension.
Taking into account the derivative operators we get the gradient expansion, where the
leading order term gives the local potential approximation (LPA).
In \eqn{effac} the operator $Z_k\equiv Z_k(\phi_x)=Z_k(\phi,p)$ is the wavefunction renormalization,
which gives the next to leading order contribution to the gradient expansion after the LPA.
When we use the latter approximation the wavefunction renormalization does not evolve, implying
that $Z_k=1$. Further terms in the derivative expansion can be
\beq\label{effacder}
\Gamma_k =\int d^d x \left[V_k(\phi_x) + \hf Z_k(\phi_x)(\partial_\mu \phi_x)^2
+ H_1(\phi_x)(\partial_\mu \phi_x)^4+H_2(\phi_x)(\Box \phi_x)^2+\ldots\right].
\eeq
If one inserts the form of the effective action in \eqn{effac} into \eqn{wettRG}, then
one obtains the following evolution equation for the potential
\beq\label{evolV}
\dot V_k = \hf\int_p \frac{\dot {\cal R}_k}{Z_k p^2+{\cal R}_k+V''_k},
\eeq
with the d-dimensional momentum integral.
Assuming that the wavefunction renormalization is momentum independent, i.e.
$Z_k(\phi,p)=Z_k(\phi)\equiv Z_k$, then one obtains the evolution equation
\bea\label{evolZ}
\dot Z_k&=&\hf\int_p \dot{\cal R}_k\biggl[
-\frac{Z_k''}{ [ p^2Z_k + {\cal R}_k + V_k'' ]^2}
+\frac{\frac2{d}Z'^2_kp^2+4Z'_k(Z'_kp^2+V'''_k)}
{(p^2 Z_k + {\cal R}_k+V_k'')^3}+\frac{\frac8{d}p^2(Z'_kp^2+V'''_k)^2
\left(Z_k+\partial_{p^2}{\cal R}_k\right)^2}
{(p^2 Z_k + {\cal R}_k+V_k'')^5}\nn
&&-2\frac{\left(Z'_kp^2+V'''_k\right)^2\left(Z_k+\partial_{p^2}{\cal R}_k
+\frac2{d}p^2\partial^2_{p^2}{\cal R}_k\right)
+\frac4{d}Z'_kp^2\left(Z'_kp^2+V'''_k\right)\left(Z_k+\partial_{p^2}{\cal R}_k\right)}
{(p^2 Z_k + {\cal R}_k+V_k'')^4}\biggr]
\eea
for the wavefunction renormalization.

If one Taylor expands the potential $V_k$ by its field variable, then one arrives at the
potential for d-dimensional one component scalar $\phi^4$ model of the form
\beq\label{dimV}
V_k = \sum_{n=1}^N \frac{g_{2n}}{(2n)!}\phi^{2n}.
\eeq
After inserting it into \eqn{evolV} then one obtains a system of ordinary differential
equations for the evolution of the couplings. By using Litim's regulator
in \eqn{regopt} the momentum integral in \eqn{evolV} can be analytically performed in any
dimensions. It is
\beq\label{litdimVdot}
\dot V_k = 2 v_d k^d \frac2{d}\frac{k^2}{k^2+V''_k},
\eeq
where
\beq
v_d = \frac1{2^{d+1}\pi^{d/2}\Gamma(d/2)},
\eeq
with $\Gamma(d)$ the Gamma function. The form of the evolution equations for the
dimensionful couplings is
\beq\label{beta}
\dot g_i = \beta_i(g_j,k),
\eeq
with the $\beta$ functions. In case of the $\phi^4$ model their general forms are
\beq
\beta_i(g_j,k) = \partial_\phi^i \left.\left(\hf\int_p \frac{\dot {\cal R}_k}
{p^2+{\cal R}_k+V''_k}\right)\right|_{\phi=0},
\eeq
in LPA, which becomes
\beq
\beta_i(g_j,k) = \partial_\phi^i \left.\left(
2 v_d k^d \frac2{d}\frac{k^2}{k^2+V''}\right)\right|_{\phi=0},
\eeq
if one uses Litim's regulator.

\subsection{Evolution equations}

One can reformulate the evolution equations in \eqn{beta} and the $\beta$ functions into
dimensionless expressions according to
\beq\label{dlbeta}
\dot{\tilde{g}}_i = -d_i \tilde g_i+\alpha_i(\tilde g_j)
\equiv \tilde \beta_i(\tilde g_j),
\eeq
where
$\alpha_i(\tilde g_j)=\beta_i(g_j k^{-d_j},1)$. The dimensionless couplings can be related to
the dimensionful ones as $\tilde{g}_i=k^{-d_j} g_i$ with $d$ the canonical (mass) dimension.

The evolution equations are usually highly nonlinear system of ordinary differential
equations, which have no analytic solutions in general. If the theory space which is
spanned by the dimensionless couplings is of high dimension, then it is extremely
difficult map out the whole phase structure numerically. Therefore it can be useful to find
such a tool which enables us not to avoid the most important parts of the theory space, i.e.
where the evolution slows down or even stops. The points where the latter situation takes
place are called fixed points. The fixed point of the evolution equations is defined as the
zeros or stationary points of the evolution equations, i.e. which satisfies
\beq\label{fp}
\dot{\tilde{g}}_i = 0.
\eeq
The couplings which are the solutions of the system of algebraic equations in \eqn{fp} are
denoted by $\tilde g_i^*$. Naturally the fixed points can be usually found only numerically.
In the vicinity of the fixed points, due to the slowing
down of the RG evolution, the flow equations can be linearized. We note that the
linearized flow equations can differ around different fixed points. If we introduce
$y_i=\tilde g_i-\tilde g_i^*$ we obtain the linearized evolution equations
\beq
\dot y_i =M_{ij}y_j,
\eeq
with the matrix
\beq\label{linM}
M_{ij}=\frac{\partial\tilde \beta_i}{\partial\tilde g_j},
\eeq
that can be constructed by taking the derivative of the $\beta$ functions w.r.t.
the dimensionless couplings. The eigenvalues of the matrix $M$ are denoted by
$s_n$. One can diagonalize $M$, with the help of a linear transformation represented
by the matrix $S$ which satisfies the relation $S_{ik}^{-1}M_{kl}S_{ln}=\delta_{in}
s_n$. We introduce $z_i=S_{ik}^{-1} y_k$ in order to decouple the linearized flow
equations according to
\beq
\dot z_i = s_i z_i.
\eeq
Its solution reads as
\beq\label{zlinsol}
z_i = z_i(k_\Lambda) e^{s_i t} = z_i(0)\left(\frac{k}{k_\Lambda}\right)^{s_i},
\eeq
where $k_\Lambda$ is some reference scale, depending on the fixed point under
investigation it can be e.g. the UV cutoff.
The real part of the eigenvalue determines whether the trajectory is attracted
or repelled by the corresponding fixed point.

\subsection{Classification of fixed points}

Let us assume that we have two dimensionless couplings, thus the theory space is 2-dimensional.
It is quite straightforward to generalize the classification for higher dimensional theory spaces,
although the structure is a bit more complicated.
It is clear that the signs of the eigenvalues determine whether we approach
or go away from the fixed point where the linearization is performed.
In this simple example case the eigenvalues are the solutions of a second order algebraic
equation, therefore in general they are complex numbers. The imaginary part of the eigenvalues
cannot alter the distance from the fixed point, altogether they make some oscillations around it.
Thus the real parts of the eigenvalues determine the types of the scalings of the couplings.

Let us denote the eigenvalues by $s_1$ and $s_2$, and consider the UV
limits, i.e. $k\to\infty$. According to the solution in \eqn{zlinsol} we have six possibilities.
\begin{enumerate}
\item
The eigenvalues are real, $s_1,s_2~\in~\mathbb{R}$ and they are negative,
$s_1,s_2<0$. Then the trajectory approaches the fixed point, and is called an
UV attractive fixed point.
\item
$s_1,s_2~\in~\mathbb{R}$ and $s_1,s_2>0$. Then the trajectory
goes away from the fixed point, and it is called a UV repulsive fixed point.
\item
$s_1,s_2~\in~\mathbb{R}$ and with opposite signs. Then there is a direction
which flows into the fixed point, and there is another one where the flow is repelled
by the fixed point. The fixed point is called a hyperbolic point or a saddle point.
\item
The eigenvalues are complex $s_1,s_2~\in~\mathbb{C}$. They are necessarily constitute
complex conjugate pairs. Here the signs of the real parts of the eigenvalues determine how the
trajectories behave. If $\Re s_1,\Re s_2<0$, then the fixed point attracts
the trajectories. The imaginary parts of the eigenvalues give some oscillation for the
trajectory. The fixed point is called a UV attractive focal point.
\item
$s_1,s_2~\in~\mathbb{C}$ and $\Re s_1,\Re s_2>0$.
Then the trajectory is repelled by the fixed point, and it is called a UV repulsive focal point.
\item
$s_1,s_2~\in~\mathbb{C}$ and the real part is zero. Then, due to the oscillation
the trajectory circulates around the fixed point along a closed trajectory. It is called an
elliptic point, which is a specific form of a limit cycle.
\end{enumerate}
Here we assumed that the eigenvalues are nonzero. If there are zero eigenvalues, then
one should calculate those terms beyond the linearized approximation in the
$\beta$ functions in \eqn{dlbeta}, which can give nontrivial contribution. We also note
that he classification according to the IR limit $k\to 0$ using \eqn{zlinsol} gives opposite
trends for the trajectories.

\subsection{Truncation and fixed points}

\begin{figure}
\begin{center}
\includegraphics[width=6cm,angle=-90]{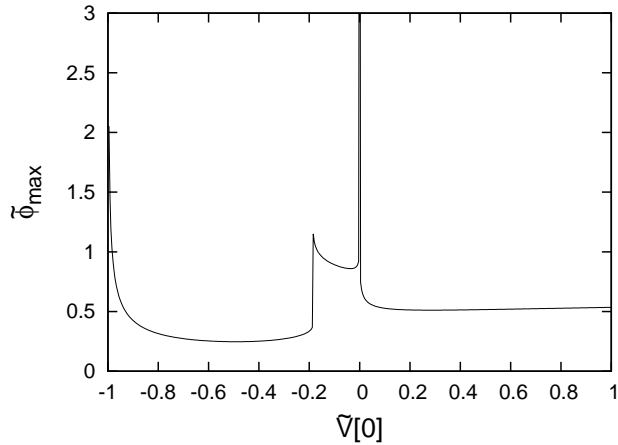}
\caption{\label{fig:whminspike} The maximal value of the field variable $\t\phi_{max}$ is presented as the
function of different initial values $\t V[0]$. Here we use the Litim IR regulator in $d=3$.
The 3 peaks suggest that there are 3 fixed points.}
\end{center}
\end{figure}

From \eqn{litdimVdot} it is easy to show that dimensionless potential $\t V_k$ satisfies
\beq\label{litrgdless}
\dot{\t V}_k = -d \t V_k +\frac{d-2}2\t \phi\partial_{\t\phi}\t V_k+ v_d \frac2{d}\frac1{1+\t V''_k}.
\eeq
The fixed point equation is
\beq\label{fpV}
\dot{\t V}^*=0,
\eeq
which provides the fixed point potential. Naturally $\t V^*$ is scale independent.
At a certain value $\t\phi_{max}$ the potential becomes singular. If $\t\phi_{max}\to\infty$
then the corresponding potential belongs to a fixed point potential
\cite{Morris:1994ki,Morris:1996kn,Morris:1996xq}.
We can solve the equation with the initial conditions
$\t V'[0] = 0$ and $\t V[0]$ which is parameterized as
\beq
\t V[0] \equiv a(x) =  -\frac{2 v_d}{d^2}\frac1{1+x}.
\eeq
We solved the differential equation for $x\in[-1,1]$ and obtained that there is
a finite value of the field variable $\t\phi_{max}$ that satisfies $\t V[\t\phi_{max}]\to\infty$.
These initial conditions correspond to fixed points.
This is demonstrated in \fig{fig:whminspike}, where the fixed points appear as the peaks
of the function. It shows that there is a fixed point at $\t V[0]=0$
which can be identified by the trivial Gaussian fixed point. At $\t V[0]\approx -0.184$ there is another
one which can be later referred to as the Wilson-Fisher fixed point. Note that a third peak builds
up as we take the limit $\t V[0]\to -1$. Although the limit makes the RG equation in \eqn{litrgdless}
singular one may suspect that a third fixed point exists, that must be an IR one.
After identifying the fixed points the form of the corresponding potentials can be determined.
The great advantage of this procedure
that we can find the fixed point potential without making any assumption about the form of it.
Naturally the existence of the fixed points should be checked by choosing other IR regulators.
It is also a relevant question how the contribution from the higher order terms in the
derivative expansion can affect the fixed points. One should keep in mind that
the most precise determination of the fixed points
and the corresponding exponents are based on this method.

It is more convenient to study the fixed points of the models by expanding the potential
according to its field variable. Then a set of algebraic equations should be solved in order to
find the fixed points and the corresponding exponents, which is mathematically much simpler to
handle than to solve the differential equation in \eqn{fpV}. If the degree of the expansion is $m$
then one expects that by increasing $m$ the exponents should converge \cite{Litim:2002cf}.
The truncations provide moderate results for the exponents, e.g. the $m$ dependence can give oscillating
results \cite{Liao:1999sh}, and their values may differ significantly from the exact ones.
Furthermore it may happen that spurious fixed points appear due to the truncation that is used
\cite{Margaritis:1987hv}. There are situations where the truncation with even $m$ finds
the Wilson-Fisher fixed point, but the odd $m$ finds only the Gaussian one. In these cases
the solution of \eqn{fpV} is needed.

Nevertheless in this article we use the truncation of the potential, furthermore we choose low numbers
of couplings. The reason is twofold. On one hand our goal is to give some introductory remarks from
the renormalization group technique and less calculations makes the method more transparent.
On the other hand the existence of the fixed points that we find is well established in most cases
in the literature. However we emphasize that reliable results requires the solution
of the fixed point potential equation in \eqn{fpV}.

\subsection{The Gaussian fixed point}

\begin{figure}
\begin{center} 
\epsfig{file=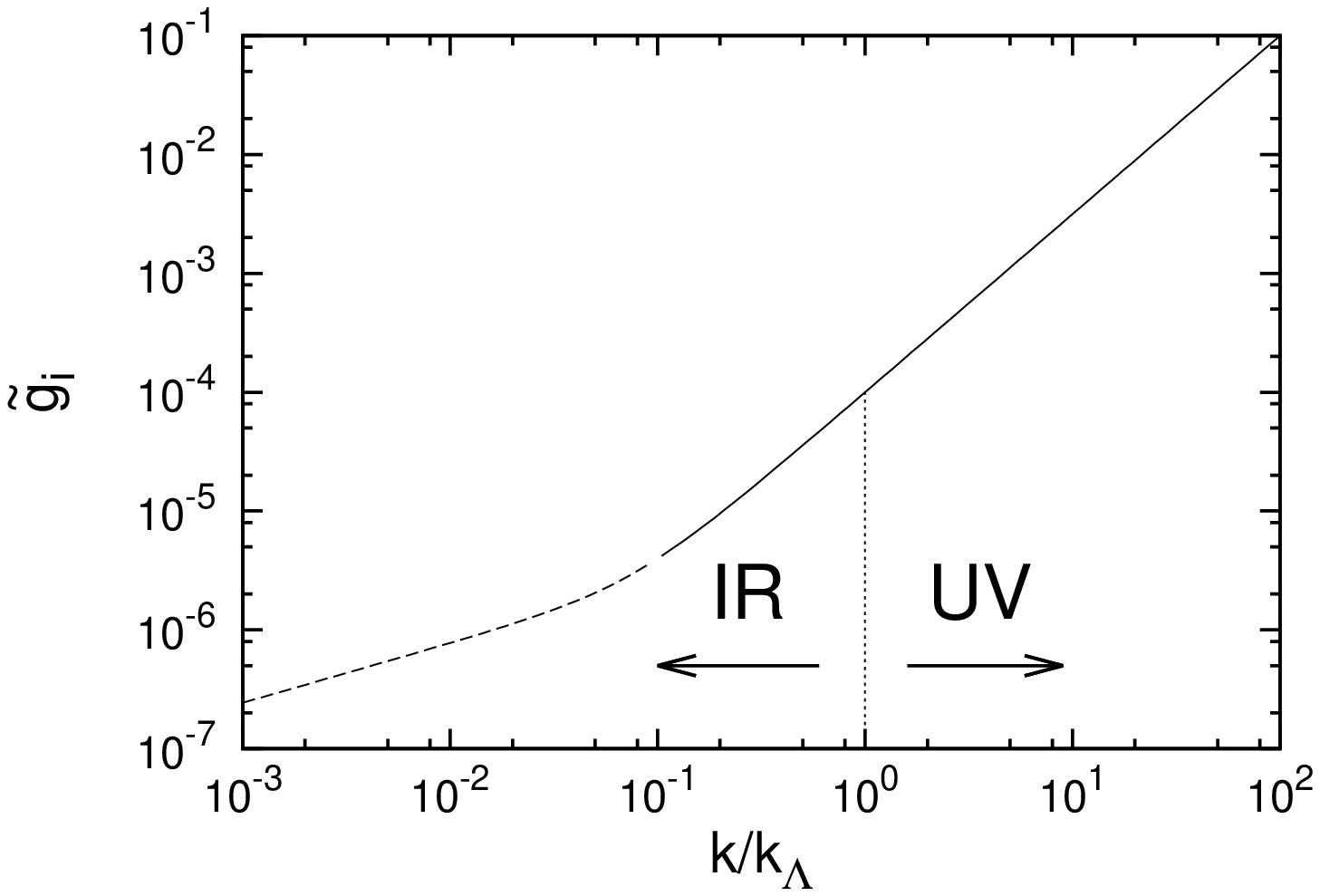,width=6cm,angle=-90}
\epsfig{file=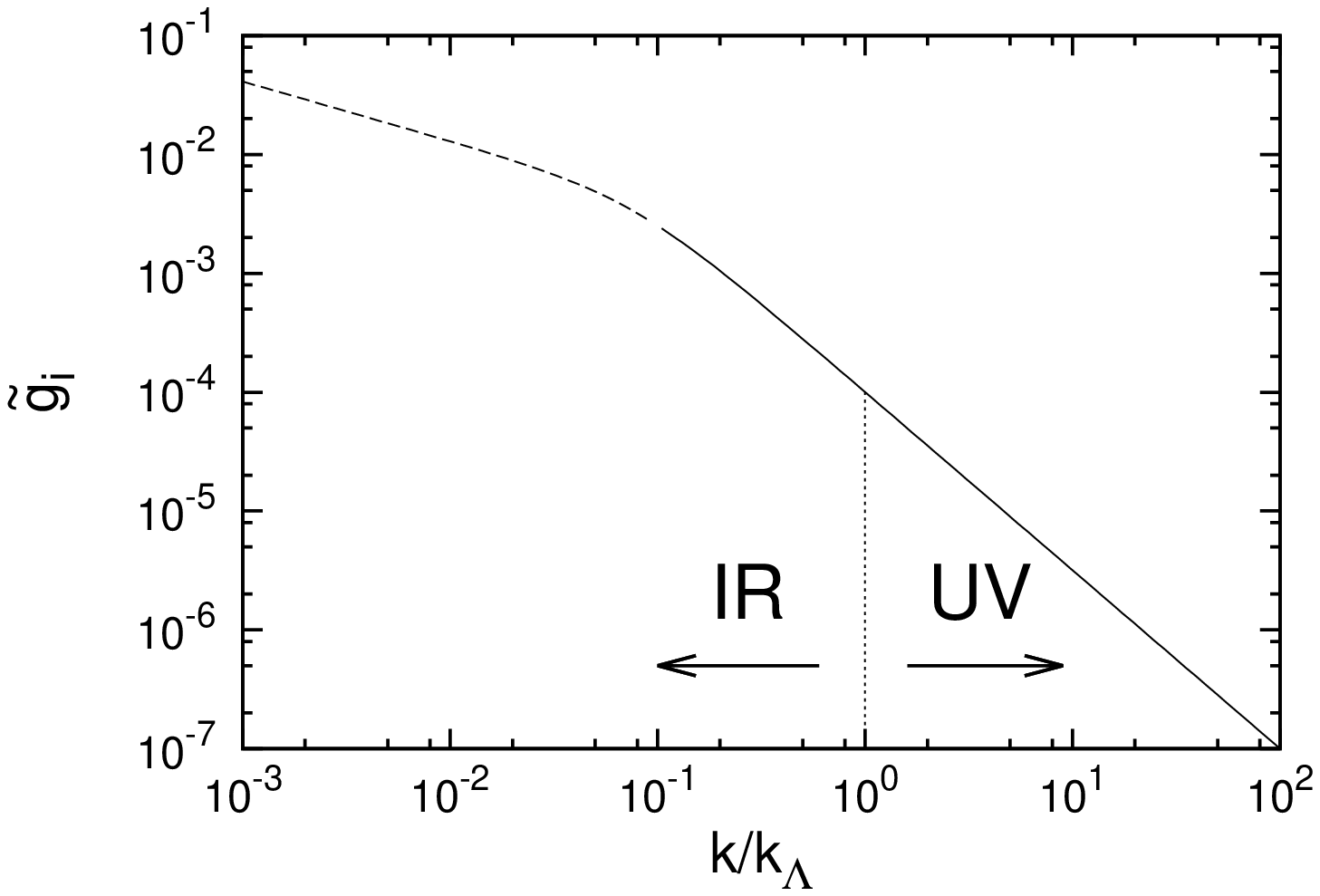,width=6cm,angle=-90}
\caption{\label{fig:irrel}
The irrelevant and relevant scalings of the couplings are shown. In the irrelevant case the
the value of the coupling tends to $\infty$ with a power law behavior as $k\to\infty$, and the
scaling remains unchanged then, since the linearization gives better and better approximation.
The value of the couplings decreases when the trajectory goes away from the UV regime. Its power
law behavior is limited to the attractive region of the fixed point, far from its scaling
regime it can deviate from the power law one. In this example it appears at $k\approx 0.1$.
Then a new linear section can appear due to a possible other fixed point in the theory space.
Naturally the new fixed point can turn the irrelevant scaling even into a relevant one.
In the other figure the relevant scaling is shown. There the $\t g\to 0$ as $k\to\infty$,
while it grows up in the opposite direction of the scale.
} 
\end{center}
\end{figure}

The GFP corresponds to the origin of the theory space, i.e.
$\tilde g_i^*=0$. It describes a free theory for massless particles.
Let us see the linearization of the flow equations in the vicinity of the GFP.
If one Taylor expands the $\beta$ functions around the origin of the theory space
then one obtains that
\beq
\tilde\beta_i = - d_i \tilde g_i + a_i\tilde g_i+a_{ijk}\tilde g_j\tilde g_k\ldots
\eeq
The matrix $M$ defined in \eqn{linM} is
\beq
M_{ij} = -d_{ij}+a_{ij},
\eeq
but one can prove that $a_{ij}=0$, when $i<j$, therefore the eigenvalues are simply equal to
the negative of the canonical dimensions, i.e.
\beq
s_i = -d_i.
\eeq
It implies that all the eigenvalues are real in the GFP.
The sign of the eigenvalue determines how the coupling behaves as we approach or go away from
the GFP. If $s_i>0$ which means that the canonical dimension
is negative $d_i<0$ then the transformed coupling $z_i\to \infty$ when the scale
$k\to\infty$ or the RG time $t$ $\to\infty$, implying that $\t g_i\to\infty$
so the trajectory is repelled by the GFP. We illustrate
it in \fig{fig:irrel}.
If the scale $k$ is lowered then the coupling tends to
zero, and it becomes less and less important as $k\to 0$, therefore we call
the coupling $\tilde g_i$ as an irrelevant one. It implies that the irrelevancy is defined
according to the IR scaling behavior, and it represents how the flow changes if the
trajectory goes away from the fixed point. Otherwise this definition
is meaningful only for the GFP, since it assumes real eigenvalues. Its generalization is however
quite straightforward by considering the real parts of the eigenvalues.
One cannot conclude from a UV behavior
to the IR one, since the trajectory can approach many other fixed points which can
totally overwrite its scaling behavior there.

The other possibility is when the eigenvalue of $M$ in \eqn{linM} is negative, i.e.
$s_i<0$ giving positive canonical dimension $d_i>0$. Then $z_i\to 0$ i.e.
$\tilde g_i\to \tilde g_i^*$ when $k$ or $t$ $\to \infty$. In this situation the
trajectory is attracted by the GFP, see \fig{fig:irrel}.
The coupling $\tilde g_i$ is said to be relevant, since
it becomes more and more important as the trajectory goes away from the fixed point.
The appearance of a new fixed point besides the GFP can overwrite the scaling properties
of the couplings. Furthermore they can save the flows from the singularities
\cite{Nagy:2010mf,Nagy:2012qz,Nagy:2012rn,Nagy:2012np}.
A coupling is said to be marginal if the corresponding eigenvalue is zero.
The dimension of the critical surface equals the number of negative eigenvalues.

We call a theory perturbatively renormalizable if $\forall ~d_i>0$ in the vicinity
of the GFP. In this case we have only renormalizable couplings.
A theory is asymptotically free if every value of the coupling tends to zero in the UV limit,
i.e. $\underset{t\to \infty}\lim {\t g_i}=0$ for $\forall ~\t g_i$
around the GFP. Examples for asymptotically free theories are e.g.  QCD, and the
3-dimensional $\phi^4$ model.

The values of the irrelevant couplings blow up in the UV, and this makes measurable
quantities nonphysical, since they will be infinitely large. However the relevant ones can
be related to the critical exponents. If $s$ is a negative eigenvalue of the matrix $M$
which corresponds to a relevant coupling, then
its negative reciprocal gives the mass critical exponent $\nu$, or the exponent of the
correlation length $\xi$, i.e.
\beq
\nu = -1/s.
\eeq
If there are several relevant couplings, which give negative eigenvalues, then the largest one can be
related to $\nu$.
We note that usually the critical exponents are identified by simple negative of the eigenvalues of $M$.
Let us recall that the scaling of the correlation length is
\beq
\xi\sim(T-T_c)^{-\nu},
\eeq
for continuous or second order phase transitions, where $T$ is the temperature, and $T_c$
is the critical temperature.

\subsection{d-dimensional $O(N)$ model}

As an example of asymptotically free theory, we treat the d-dimensional $O(N)$ model
by functional RG method. We map out its phase structure, find the fixed points and
the corresponding exponents. Although most scalar models are only toy models, but the $O(N)$
model has experimental realization for certain values of $N$, i.e. they can characterize the
following physical systems:
\begin{itemize}
\item[$N=0$] polymers,
\item[$N=1$] liquid-vapor transition, or uniaxial (Ising) ferromagnets,
\item[$N=2$] $He^2$ superfluid phase transition,
\item[$N=3$] Heisenberg ferromagnets,
\item[$N=4$] chiral phase transition for two quark flavors.
\end{itemize}

\begin{figure}
\begin{center}
\includegraphics[width=6cm,angle=-90]{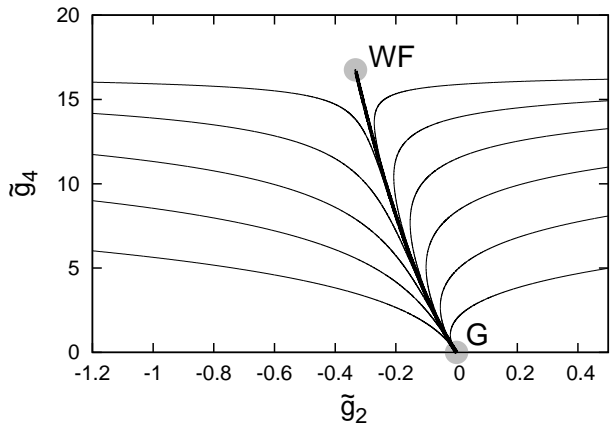}
\caption{\label{fig:ispt}
The phase structure of the 3d $\phi^4$ model treated perturbatively. The model has
a GFP and a WF fixed point. The thick line represents the separatrix. The trajectories start from
the vicinity of the GFP.
} 
\end{center}
\end{figure}

The 3d $O(1)$ or 3d $\phi^4$ model is a widely investigated model, and it possesses a nontrivial
fixed point. We use the power law regulator in \eqn{regpow} for $b=1$ since the momentum integral
in \eqn{evolV} can be analytically performed in $d=3$, similarly to Litim's one in \eqn{litdimVdot}.
After the loop integration the dimensionful potential becomes
\beq
\dot V = -\frac{k^2}{4\pi}\sqrt{k^2+V''},
\eeq
up to a field independent divergent term, which can be omitted. If the functional form of the potential in
\eqn{dimV} is inserted to the evolution equation, then one can obtain the
$\beta$ functions for the couplings. Their dimensionless forms are
\bea\label{phi4beta}
\t\beta_2 &=& -2\tilde g_2-\frac{\tilde g_4}{8\pi(1+\tilde g_2)^{1/2}},\nn
\t \beta_4 &=& -\tilde g_4+\frac{3\tilde g_4^2}{16\pi(1+\tilde g_2)^{3/2}},
\eea
where we neglected the evolution of further couplings which are generated by
the RG procedure. In the symmetric phase their effects are really negligible, but they play
crucial role in the evolution within the broken phase.
Other regulators would give qualitatively similar equations but e.g.
with different multiplication factors, or different exponents in the denominator
in the second term on the r.h.s.. Here we do not derive the perturbative RG equations, instead
we expand the $\beta$ functions in \eqn{phi4beta} by the coupling $\t g_4$. The zeroth
order approximation gives the flows driven by the canonical dimensions. They are
\beq
\t g_i = \t g_i(k_\Lambda) e^{-d_i t},
\eeq
with $g_i(k_\Lambda)$ the initial value and
\beq
d_i = d+i\left(1-\frac{d}2\right),
\eeq
the canonical dimension of the coupling. This approximation of the flow equations has a
GFP. The matrix $M$ in \eqn{linM} is
\beq
M = \begin{pmatrix}
\partial_{\t g_2}\beta_2 & \partial_{\t g_4}\beta_2\\
\partial_{\t g_2}\beta_4 & \partial_{\t g_4}\beta_4\\
\end{pmatrix}
=
\begin{pmatrix}
-2 & 0\\
0 & -1\\
\end{pmatrix},
\eeq
with the eigenvalues $s_1=-2$ and $s_2=-1$. Thus the GFP is a UV attractive point,
and both couplings are relevant. The eigenvalues and the type of the GFP do not change
if one considers further terms in the approximation. Taking into account the next term in the
expansion of the flow equations in \eqn{phi4beta} one gets
\bea
\dot{\tilde g}_2 &=& -2\tilde g_2-\frac{\tilde g_4}{8\pi}+\ord{\tilde g_i^2},\nn
\dot{\tilde g}_4 &=& -\tilde g_4+\frac{3\tilde g_4^2}{16\pi}+\ord{\tilde g_i^3}.
\eea
The structure of the theory space is plotted in \fig{fig:ispt}.
The model has two fixed points. They can be found by solving the system of algebraic equations
$\t\beta_2=0$ and $\t\beta_4=0$. The first is the GFP, $\t g_2^*=\t g_4^*=0$, with the
eigenvalues $s_1=-2$ and $s_2=-1$ found at the tree level previously.
Furthermore, there is a nontrivial fixed point at $\t g_2^*=-1/3$ and $\t g_4^*=16\pi/3$. The
matrix $M$ now has the form
\beq
M=\begin{pmatrix}
-2 & -\frac1{8\pi}\\
0 & -1+\frac{3\t g_4}{8\pi}\\
\end{pmatrix}_{g_2^*=-1/3,\t g_4^*=16\pi/3}
=\begin{pmatrix}
-2 & -\frac1{8\pi}\\
0 & 1\\
\end{pmatrix}
\eeq
Its eigenvalues are $s_1=-2$ and $s_2=1$, so the fixed point is a saddle point
or a hyperbolic point. It is called Wilson-Fisher (WF) fixed point. It appears in $\phi^4$
models with dimension $2<d<4$. When $d\to 4$ then the WF fixed point tends to the origin
and in $d=4$ it melts into the GFP. Since the critical exponent $\nu$
of the correlation length $\xi$ is identified as the negative reciprocal of the single
negative eigenvalue of the matrix $M$ coming from the linearization of the evolution equations,
then it gives $\nu=-1/s_1=1/2$. The approximation of the
model with two couplings makes the problem a mean field type one.

The 3d $\phi^4$ model has two phases. The trajectories tending right from the WF fixed point
in \fig{fig:ispt} correspond to the symmetric phase. The trajectories tending left belong to the
broken phase. There the $Z_2$ symmetry ($\phi\leftrightarrow -\phi$) is broken.

\begin{figure}
\begin{center}
\includegraphics[width=6cm,angle=-90]{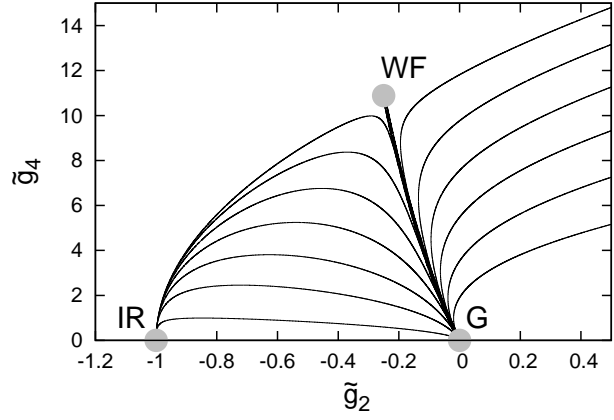}
\caption{\label{fig:isrg}
The phase structure of the 3d $\phi^4$ model calculated from the flow equations
in \eqn{phi4beta}. The model has a GFP, a WF and an IR fixed point. The latter one corresponds
to a crossover fixed point between the UV and the IR ones.
The thick line represents the separatrix. The trajectories start from
the vicinity of the GFP.
} 
\end{center}
\end{figure}

Further terms in the expansion in $\t g_4$ do not give further fixed points, but
only changes qualitatively the position of the WF fixed point. If one considers the exact
RG flow equations in the two couplings in \eqn{phi4beta} we have the GFP at the origin and the 
WF fixed point at $\t g_2^*=-1/4$ and $\t g_4^*=2\sqrt{3}\pi$. The matrix $M$ is
\beq
\begin{pmatrix}
-2 +\frac{\t g_4}{16\pi(1+\t g_2)^{3/2}}& -\frac1{8\pi(1+\t g_2)^{1/2}}\\
-\frac{9\t g_4^2}{32\pi(1+\t g_2)^{5/2}} & -1+\frac{3\t g_4}{8\pi(1+\t g_2)^{3/2}}\\
\end{pmatrix}_{g_2^*=-1/4,\t g_4^*=2\sqrt{3}\pi}
=\begin{pmatrix}
-\frac53 & -\frac1{4\sqrt{3}\pi}\\
-4\sqrt{3}\pi & 1\\
\end{pmatrix}.
\eeq
Its eigenvalues are $s_1=-2$ and $s_2=4/3$. The WF fixed point remains
a saddle point. The higher order terms in approximations (e.g. consideration of
further couplings, inclusion wavefunction renormalization or even higher order terms
in the derivative expansion) cannot change the type of the fixed point only the position
is shifted further. We should emphasize that although the qualitative picture of the phase
structure can be nicely understood within this approximation, but the parameterization in \eqn{dimV}
is not suitable to study the WF fixed point. There one should use either an expansion around the
nontrivial minimum of the potential or no expansion.
We illustrate the phase structure in \fig{fig:isrg}.
Interestingly the trajectories tend to a single point in the broken phase. In might assume
that there can be a further fixed point in the IR limit. That seems strange
according to the $\beta$ functions in \eqn{phi4beta}, since we can find all the fixed
point solutions analytically. The only possibility is that the ratios in \eqn{phi4beta}
can tend to zero in the way that both the numerator and the denominator vanish. If they
are of the form of 0/0, then it implies that the values of the couplings are
$\t g_2^*=-1$ and $\t g_4^*=0$. It corresponds to an
effective potential of the form $\t V_0=-\phi^2/2$. Clearly this fixed point cannot
be reached by any expansions around the GFP since this point is a singular point of the
evolution equations.

The effective potential of the symmetric phase can be handled
quite easily numerically in contrast to the one in the broken phase, where one can face
a very difficult numerical and theoretic problem \cite{Berges:2000ew}.
The reason is that there is a condensate in the broken phase
\cite{Alexandre:1998ts,Polonyi:2005cq,Nagy:2006pq,Polonyi:2010yt,Alexandre:2012hn,
Alexandre:2012ht}. It can be considered as a macroscopic object constituting
of a huge amount of soft modes. Its measure gives a dynamically appearing momentum scale
in the model, which implies that there cannot be such modes which momentum is that is larger than
the one characterizing the condensate \cite{Wetterich:1989xg,Boyanovsky:1998yp,
Alexandre:1998ts}. The evolution stops when one reaches the scale
of the condensate, which manifests in the form that the flows arrive at the singularity.
Is there really a singularity there, or it is only a numerical artifact? It is known that
the Wetterich equation in \eqn{wettRG} should not drive the flows into singularity. However
the approximation due to the strong truncation of the functional ansatz for the
effective action, or in the derivative expansion may induce a singular behavior suggesting
that the singularity is a numerical artifact. Therefore
it is argued that the singularity can be avoided by a proper choice of the IR regulator, and
the flow can reach arbitrarily small scales. However it was shown for a huge set of
scalar models \cite{Nagy:2010mf,Nagy:2012qz,Nagy:2012rn,Nagy:2012np}
that such a singularity possesses a specific scaling behavior induced by the IR
fixed point, therefore it may have a significant physical importance.
The singularity in the RG evolution is always reached, but the RG equations do not loose their
validity, they simply stop at a finite scale \cite{Pangon:2009wk,Pangon:2009pj}.
This finite scale also appears when it seems that the evolution avoids the singularity
with special regulators. In those cases, after an abrupt change in the
scaling of the couplings, a marginal scaling appears towards the deep IR regime without any
universality. Therefore the dynamically appearing finite momentum scale seems to be the only
universal behavior in the IR.

The dynamical scale is induced by an IR fixed point in the broken phase. One can
determine numerically the critical exponents in its vicinity. The
IR fixed point can be uncovered analytically from the $\beta$ function in \eqn{phi4beta}
by reparameterization of the couplings
according to $\omega=1+\t g_2$, $\chi=\t g_4/\omega$ and $\partial_\tau=\omega \partial_t$.
We obtain that
\bea\label{omch}
\partial_\tau \omega&=& 2\omega(1-\omega)-\frac{\chi \omega}{8\pi},\nn
\partial_\tau \chi &=& -\chi+\frac{\chi^2}{4\pi}.
\eea
The reparameterized flow equations give the Gaussian ($\omega^*_G=1$,
$\chi^*_G=0$), and the WF ($\omega^*_{WF}=3/4$, $\chi^*_{WF}=4\pi$) fixed points, however
another one appears at $\omega^*_{IR}=0$ and $\chi^*_{IR}=4\pi$. The latter can be identified
with the IR fixed point where the trajectories of the broken phase meet.
The corresponding eigenvalues are $s_1=1$ and $s_2=3/2$ expressing the
UV repulsive (IR attractive) nature of the IR fixed point.

\begin{figure}
\begin{center} 
\epsfig{file=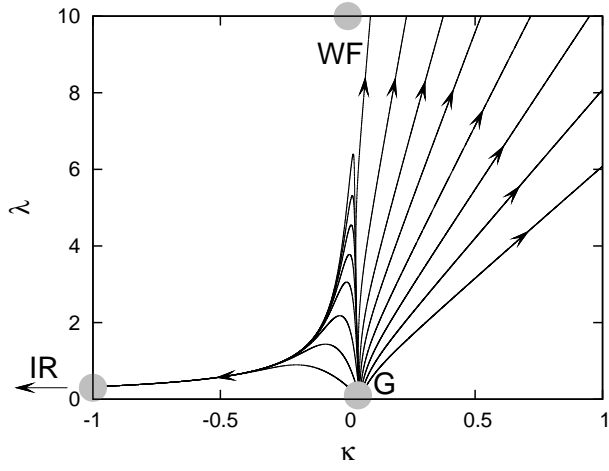,width=6cm,angle=-90}
\caption{\label{fig:onphase} The theory space of the 3d $O(N)$ model is shown.
The flows belonging to the symmetric (broken symmetric) phase tend to right (left),
respectively. The trajectories start from the vicinity of the GFP.
Again, the Wilson-Fisher fixed point plays the role of the crossover fixed point
between the UV (which is now the Gaussian) and the IR ones.
} 
\end{center}
\end{figure}

The determination of the critical exponents in the vicinity of the IR fixed point is
demonstrated in the framework of the d-dimensional $O(N)$ model. The potential has the form
\beq\label{pot}
\tilde V = \sum_{n=2}^{N_\lambda}\frac{\lambda_n}{n!}(\rho-\kappa)^n,
\eeq
with $N_\lambda$ the degree of the Taylor expansion and the dimensionless
couplings $\kappa$ and $\lambda_n$ for $n\ge 2$. For shorthand we use $\lambda_2=\lambda$.
The further dimensionless quantities are denoted by $\sim$. The introduction
of the coupling $\kappa$ serves a better convergence in the broken symmetric
phase. The evolution equation for the potential reads as
\bea\label{potev}
k\partial_k\tV &=& -d \tV+(d-2+\eta)\t\rho\tV'+\frac{4v_d}{d}\left(1-\frac{\eta}{d+2}\right)
\left(\frac1{1+\tV'+2\t\rho\tV''}+\frac{N-1}{1+\t V'}\right),
\eea
by using Litim's regulator, with the notation $'=\delta/\delta\rho$.
In \eqn{potev} we introduced the anomalous dimension $\eta$ which is defined via the
wavefunction renormalization according to $\eta=-d\log Z/d\log k$ and can be calculated by
means of the couplings as
\beq\label{etak}
\eta = \frac{16 v_d}{d}\frac{\kappa\lambda^2}{1+2\kappa\lambda}.
\eeq
The inclusion of $\eta$ in the RG equation mimics
the evolution of the wavefunction renormalization. We note that a more
precise treatment can be obtained if one Taylor expand the evolution
equation for $Z$ in \eqn{evolZ}.
Again, we start with the 3d $O(1)$ model. It is instructive to retreat the phase structure,
since the IR fixed point moves to infinity. Now its flow equations are
\bea\label{klflow}
\dot \kappa &=& -\kappa+\frac1{2\pi^2(1+2\kappa\lambda)^2},\nn
\dot \lambda &=& -\lambda+\frac{3\lambda^2}{\pi^2(1+2\kappa\lambda)^3},
\eea
for the first two couplings if we set $\eta=0$ and $\lambda_n=0$ for $n> 2$.
The $O(N)$ model in $d=3$ has two phases. The typical phase structure is depicted in
\fig{fig:onphase} for the couplings in \eqn{klflow}.
From the flow equations in \eqn{klflow} one can find two fixed points in the model.
The Gaussian fixed point can be found at $\kappa_G^*=1/2\pi^2$ and $\lambda_G^*=0$,
although it is not situated in the origin due to the redefinition of the field variable.
The linearization of the flow equations
around the GFP gives a matrix with negative eigenvalues ($s_{G1}=-1$ and
$s_{G2}=-1$), i.e. the fixed point is repulsive or UV attractive.
The WF fixed point can be found at the values
of the couplings $\kappa^*_{WF}=2/9\pi^2$ and $\lambda^*_{WF}=9\pi^2/8$, with eigenvalues
$s_{WF1}=1/3$ and $s_{WF2}=-2$, so it is a hyperbolic point.

We usually identify the critical exponent $\nu$ of the correlation length $\xi$
in the vicinity of the WF fixed point by taking
the negative reciprocal of the single negative eigenvalue, which
gives $\eta_{WF}=1/2$ in this case. Let us notice
that the flows tend to a single curve beyond the WF fixed point in the broken phase
into the IR fixed point at $\kappa_{IR}^*\to-\infty$ and $\lambda_{IR}^*=0$.
Again, with the help of a rescaling of the couplings the attractive IR
fixed point can be uncovered and one finds the following pair of evolution equations
\bea
\partial_\tau \omega &=& 2\omega(1-\omega)-\frac{\ell\omega}{\pi^2}(3-4\omega),\nn
\partial_\tau \ell &=& \ell(5\omega-6)+\frac{9\ell^2}{\pi^2}(1-\omega),
\eea
where $\omega=1+2\kappa\lambda$, $\ell=\lambda/\omega^3$ and $\partial_\tau=k\partial_k/\omega$.
The static equations now have the Gaussian ($\ell_G^*=0$, $\omega_G^*=1$),
the WF ($\ell_{WF}^*=\pi^2/3$, $\omega_{WF}^*=3/2$) and the IR
($\ell_{IR}^*=2\pi^2/3$, $\omega_{IR}^*=0$) fixed point solutions. Naturally the Gaussian and the
WF ones has the same behavior as was obtained from direct calculations. However the new IR
fixed point indeed corresponds to the values $\kappa_{IR}^*\to-\infty$ and $\lambda_{IR}^*=0$,
and the linearization in its vicinity gives the eigenvalues $s'_{IR1}=6$ and $s'_{IR2}=0$,
a positive and a zero one, showing that
the fixed point is IR attractive, in accordance with the flows in \fig{fig:onphase}.

\begin{figure}
\begin{center} 
\epsfig{file=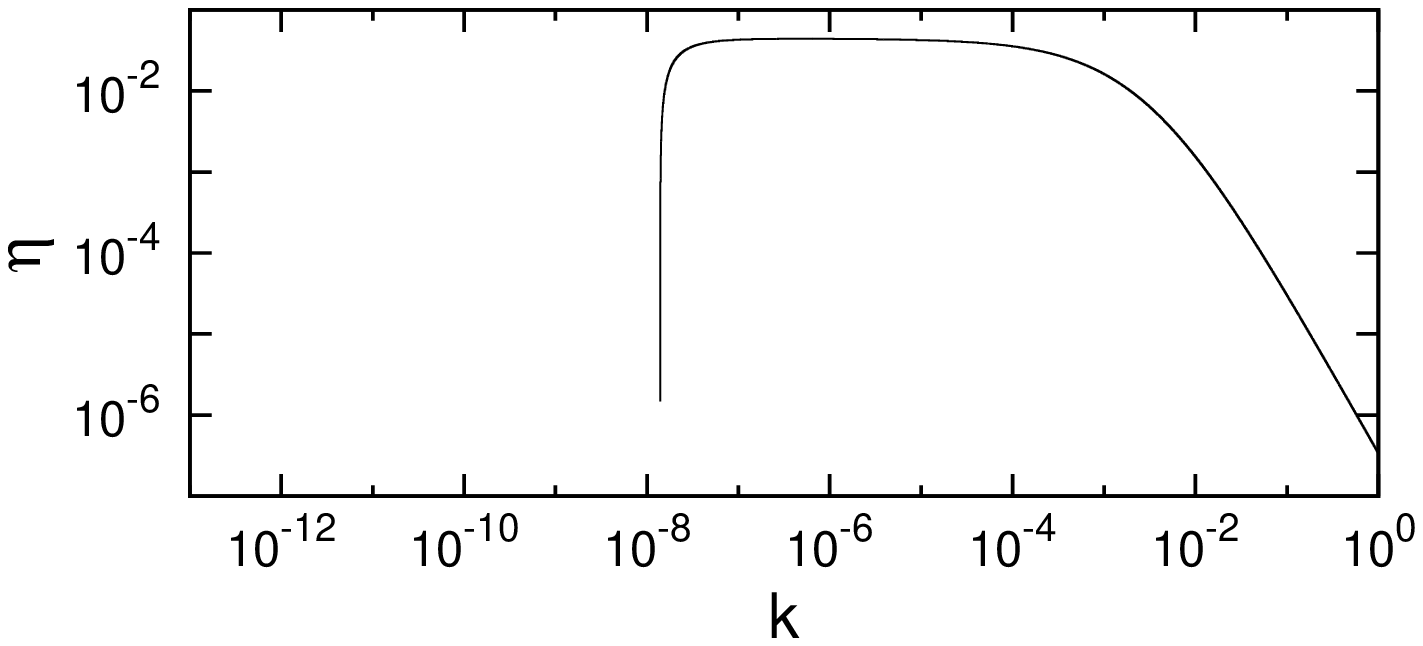,width=4cm,angle=-90}
\epsfig{file=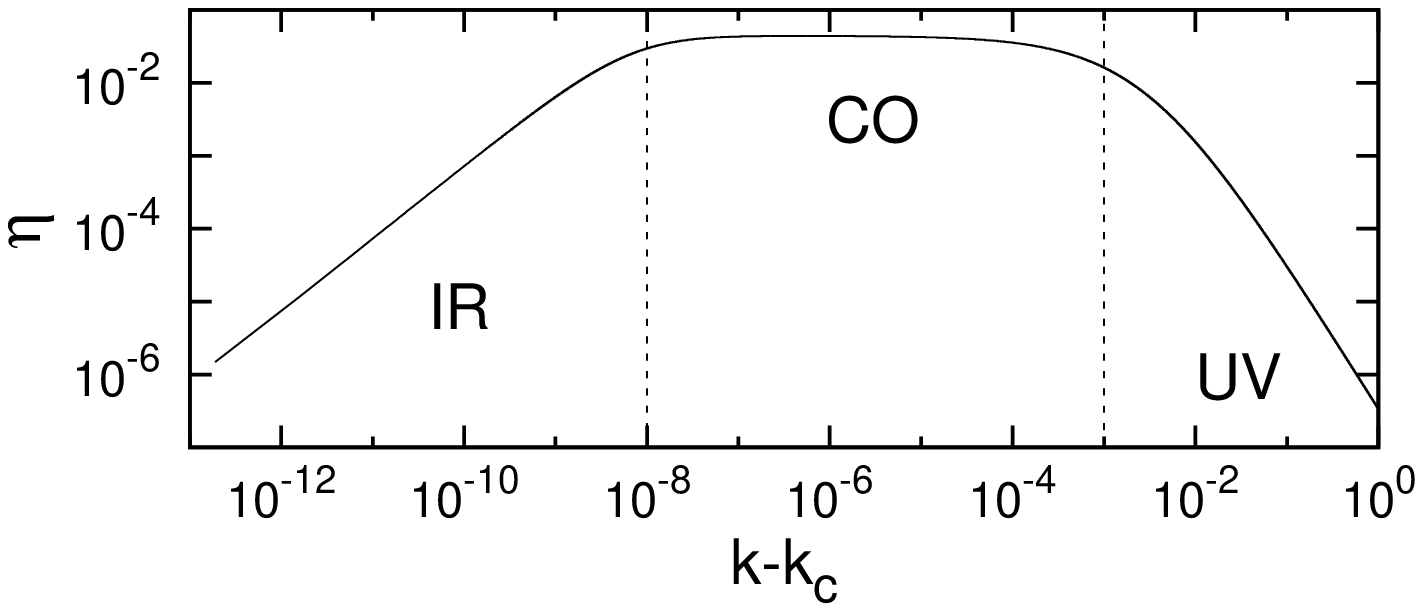,width=4cm,angle=-90}
\caption{\label{fig:oneta} The evolution of the anomalous dimension $\eta$ is presented.
In the left figure the flow of $\eta$ shows a strong singularity as the function of the
scale $k$, while this singularity becomes a power law like convergence as the function
of the shifted scale $k-k_c$ in the right one.}
\end{center}
\end{figure}

If we let many couplings and $\eta$ evolve, then the theory space does not change
significantly, but the IR fixed point can be observed more easily.
In order to demonstrate it we plotted the flow of the anomalous dimension in \fig{fig:oneta}.
We also plotted $\eta$ as the function of the shifted scale $k-k_c$ in the
right figure of \fig{fig:oneta}. In the UV region the Gaussian fixed point drives the evolution
of the anomalous dimension.  In this regime it grows according to the power law scaling
$\eta_{UV}\sim (k-k_c)^{-2}$. There is a crossover (CO) scaling region between
$10^{-8}\lesssim k-k_c\lesssim 10^{-4}$ where a plateau appears giving a constant
value for $\eta_{CO}\approx 0.043$ due to the WF fixed point. Going further in the evolution
towards the smaller values of $k$ below $k-k_c\sim 10^{-8}$ one can find a third scaling regime.
It appears a simple singularity in the left figure, but the shifted scale $k-k_c$ clearly
uncovers the power law scaling of the anomalous dimension there according to
$\eta_{IR}\sim(k-k_c)^1$. This scaling region is induced by the IR fixed point.

The evolution of the other couplings also shows such type of scaling regimes with
similar singularity structure in the IR limit. There the power law
scaling behaviors also take place as the function of $k-k_c$ with the corresponding exponents.
This suggests that the appearing singularities are not artifacts and the RG flows can be
traced till the value of $k_c$.

We note that one can find such a value of $N_\lambda$ where the evolution does not stop as
in the upper figure in \fig{fig:oneta}, but after a sharp fall during the flow of $\eta$ it
continues its RG evolution marginally giving a tiny value there. However, the singular-like fall
possesses the same power-law like behavior as the function of $k-k_c$, for any value of
$N_\lambda$. It suggests that the value of $\eta$ rapidly falls to zero at $k_c$ and it is due to
the numerical inaccuracy, whether the RG evolution survives the falling and can be traced to
any value of $k$, or the flows stop due to the appearing singularity.
It strongly suggests that the singular behavior with its uncovered IR scaling
for the shifted scale $k-k_c$ is not an artifact but is of physical importance, since
the value of the scale $k_c$, which appears there can characterize the condensate as the smallest
available scale. If one tunes the initial values of the couplings
$\kappa_{k_\Lambda}$, $\lambda_{k_\Lambda}$ etc., to the
separatrix then the value of $k_c$ decreases. It enables us to define the correlation
length $\xi$ in the IR scaling regime as the reciprocal of the scale where the evolution stops,
i.e. $\xi=1/k_c$. As the initial couplings approach the separatrix the stopping scale
$k_c\to 0$ therefore the correlation length diverges. Naturally it is infinite in the
symmetric phase. We can identify the reduced temperature $t$
in the $O(N)$ model as the deviation of the UV coupling $\kappa_{k_\Lambda}$ to its
UV critical value, i.e. $t\sim \kappa^c_{k_\Lambda}-\kappa_{k_\Lambda}$.
By starting evolutions for different values of the UV coupling $\kappa_{k_\Lambda}$
then one can get the corresponding values $\xi$. One should fine tune the initial coupling
to get larger and larger values of $\xi$ to reach the IR scaling regime. The critical
initial value of $\kappa^c_{k_\Lambda}$ can be determined by the trick, where one should fine
tune its value in the log-log plot of the $t,\xi$ plane till one obtains a straight line there.
This holds for continuous phase transition. The absolute value of the negative slope of the line
provides us the exponent $\nu$ corresponding to the correlation length.

\begin{figure}
\begin{center} 
\epsfig{file=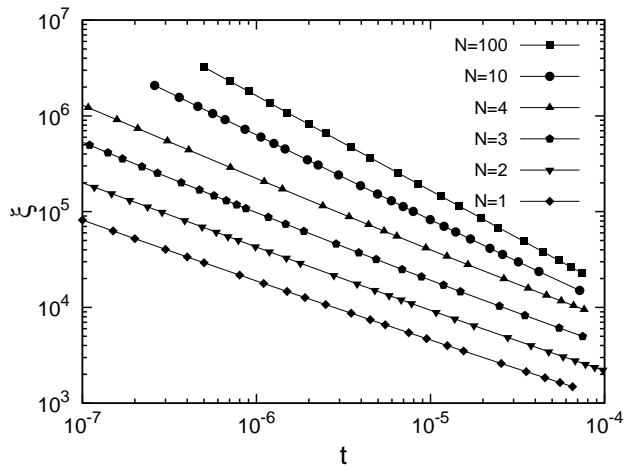,width=6cm,angle=-90}
\caption{\label{fig:d3on} The scaling of the correlation length
$\xi$ as the function of the reduced temperature $t$, for various values of $N$.
} 
\end{center}
\end{figure}

In the 3d $O(N)$ model a second order or continuous phase
transition appears according to the power law scaling
\beq
\xi\sim t^{-\nu}.
\eeq
The results are plotted in \fig{fig:d3on}.
For a given value of $N$ we obtain power-law like behavior for the scaling
of $\xi$, and the slope of the line provides the exponent $\nu$ in the log-log scale.
The obtained results are listed in \tab{tab:nu}. We denoted the WF (IR) values
of $\nu$ as $\nu_{\mbox{WF}}$ ($\nu_{\mbox{IR}}$), respectively.
\begin{table}
\begin{center}
\begin{tabular}{|c||c|c|c|c|c|c|}
\hline
  $N$ & 1 & 2 & 3 & 4 & 10 & 100 \\
\hline
 $\nu_{\mbox{IR}}$ & 0.624 & 0.666 & 0.715 & 0.760 & 0.883 & 0.990 \\
\hline
 $\nu_{\mbox{WF}}$ & 0.631 & 0.666 & 0.704 & 0.739 & 0.881 & 0.990 \\
\hline
\end{tabular}
\end{center}
\caption{\label{tab:nu} The critical exponent $\nu$ in the $O(N)$ model
for various values of $N$.}
\end{table}
The results show high coincidence. The values $\nu_{\mbox{WF}}$
are taken from results obtained from derivative expansion up to the second
order, since this approximation is the closest to our treatment.
One can conclude that the exponent $\nu$ can be also determined from the
scaling around the IR fixed point, and has the same value as was obtained
around the WF fixed point.

In arbitrary dimensions one can consider the evolution of the couplings $\kappa$ and $\lambda$
around the GFP. It is situated at
$\kappa^*_G=2^{1-d}\pi^{-d/2}(N+2)/d(d-2)\Gamma(d/2)$ and $\lambda^*_G=0$. The corresponding
matrix $M$ is
\beq
M=\begin{pmatrix}
2-d & \frac{3(N+2)4^{1-d}\pi^{-d}}{d-2\Gamma(1+d/2)^2}\\
0 & d-4\\
\end{pmatrix},
\eeq
therefore the eigenvalues are $s_{G1}=2-d$ and $s_{G2}=d-4$. When $2<d<4$, then the GFP
is UV attractive and the model is asymptotically free. However one of the eigenvalues become
zero at $d=4$, which implies that we should go beyond the linear approximation in order
to determine the scaling behavior of the corresponding coupling $\lambda$. If we consider
the 4-dimensional version of \eqn{klflow}, then it turns out that the quadratic term in
$\lambda$ appears with a positive multiplicative constant $3/4\pi^2$, therefore the 
UV evolution becomes irrelevant. Naturally the relevant scaling of $\kappa$ remains
unchanged. In $d=4$ the Gaussian fixed point behaves as a saddle point,
which can bring the value of $\lambda$ to infinity, so the model is not asymptotically free
anymore and it becomes non-renormalizable. Strictly speaking the model is
perturbatively renormalizable, since there is only one marginal coupling in the
theory space, therefore the perturbative expansion can be made finite without
introducing new vertices in every order. The non-renormalizability appears as a consequence
of the marginal irrelevance of the coupling $\lambda$. It can be avoided only by setting the initial value
of $\lambda_{k_\Lambda}$ to zero. This is known as the triviality problem of the 4d $\phi^4$ model.
We note that in the Standard Model the triviality of the Higgs sector can be avoided by the inclusion
of further couplings to the matter and the bosonic fields. The extended theory space
can contain further fixed points and they can overwrite UV properties of the GFP.
A similar divergence appears in QED \cite{Gockeler:1997dn,Gies:2004hy}, where a singularity of
the coupling appears at a large finite scale $k$, and is called the Landau pole.

\section{Asymptotic safety}\label{sect:as}

The concept of perturbative renormalizability and the asymptotic freedom is restricted
to the GFP. It guarantees that the physical quantities which are calculated from the model
do not suffer from divergences. However if there are irrelevant couplings
in the GFP which are crucial in describing the model that is investigated,
(e.g. the Newton constant in QEG, which makes the GFP a hyperbolic one) then the
systematic removal of the divergences may even induce infinitely many new important couplings
which implies that the model disables us to give any physical predictions.

However there can be further nontrivial (non-Gaussian) fixed points (NGFP),
where the physically important couplings are relevant, which means that the theory can give
finite physical quantities. This is the basic idea of the asymptotic safety.
In general sense the asymptotic safety means that the theory is free
from divergences if the cutoff is removed to infinity assuming that the corresponding
fixed point possesses a finite number of UV attractive directions.
In the fixed point those couplings should scale in relevant manner which are crucial to obtain
finite physical quantities. Naturally there can be certain couplings which are generated by
the RG procedure, but are unimportant to describe the physical process at that energy scale.
It is necessary to have finite number of relevant couplings, otherwise every trajectory
would tend to the fixed point and the theory would not be predictive. The definition of 
the asymptotic safety is not restricted to the GFP. It implies that there is a UV NGFP in the
theory space. The asymptotic safety requires that the eigenvalues corresponding to the
linearized RG flows of the physically important couplings around the UV NGFP should have
negative real parts.

Examples for asymptotically safe theories are e.g. the Gross-Neveu model,
the nonlinear $\sigma$ model with dimension $2<d<4$, the 2d sine-Gordon model, and it seems
that the quantum Einstein gravity also shows asymptotic safety.

\subsection{The nonlinear $\sigma$ model}

The nonlinear $\sigma$ model (NLSM) in general describes the dynamics of a map $\varphi$
from a d-dimensional manifold ${\cal M}$ to a N-dimensional manifold ${\cal N}$.
The model is renormalizable in $d=2$ and is asymptotically free
\cite{Brezin:1975sq,Brezin:1976ap}.
However if one goes beyond $d=2$ then it becomes nonrenormalizable since the
existing UV attractive Gaussian fixed point becomes a hyperbolic one.
However a nontrivial UV fixed point arises \cite{Brezin:1975sq}, which saves
the UV limit, and the model becomes asymptotically safe.
The action of the model contains only derivative interactions, e.g.
\beq\label{nlsm_action}
S = \hf \zeta \int d^d x\partial_\mu\varphi^\alpha\partial^\mu\varphi^\beta
h_{\alpha\beta}(\varphi)
\eeq
where $h_{\alpha\beta}$ is the dimensionless metric, $\zeta=1/g_0^2$ and its dimension is
$[g_0]=k^{(2-d)/2}$.
The RG equations can be derived by background field method. The perturbative RG flow gives
\cite{Codello:2008qq}
\beq\label{nlsmrgpt}
\beta_{\t g_0} = \frac{d-2}2 \t g_0 -c_d\frac{R}{N}\t g_0^3,
\eeq
where $R$ is the Ricci scalar, and
\beq
c_d=\frac1{(4\pi)^{d/2}\Gamma(d/2+1)}.
\eeq
There are two fixed points of the flow equation in \eqn{nlsmrgpt}. At $g^*_{0G}=0$ we have a GFP,
with the eigenvalue $s_G=(d-2)/2$. It is clearly positive for $d>2$ giving irrelevant
scalings and divergences in the UV limit. The other fixed point is situated at
$g^{*2}_{0UV}=(d-2)N/(2c_d N)$ giving the eigenvalue $s_{UV}=2-d$, which induces relevant
scaling, therefore the NLSM exhibits asymptotic safety, yet at perturbative level.
The critical exponent $\nu$ of the correlation length $\xi$ equals the negative reciprocal of
$s_{UV}$ giving $\nu=1/(d-2)$.

The functional RG flow equation results in the equation \cite{Codello:2008qq}
\beq
\beta_{\t g_0} = \frac{d-2}2 \t g_0
-\frac{c_d\frac{R}{N}\t g_0^3}{1-2c_d\frac{R}{N(d+2)}\t g_0^2}.
\eeq
Again, we have two fixed points. The origin is a GFP with the same eigenvalue $s_G=(d-2)/2$,
and we have a UV NGFP at $g^{*2}_{0UV}=N(d^2-4)/(4c_d d R)$. It is a UV attractive point
with eigenvalue $s_{UV}=-2d(d-2)/(d+2)$. We note that this value is smaller then the one
that was obtained perturbatively, therefore the exponent $\nu$ is greater, e.g.
it is $\nu=5/6$, when $d=3$.
In the NLSM the perturbative and the exact RG flow equations give qualitatively similar 
results, and the functional RG method changes only the position of the fixed points, and
the concrete value of exponents. The asymptotic safety appears in both the perturbative
and the exact RG treatments.

\begin{figure}
\begin{center} 
\epsfig{file=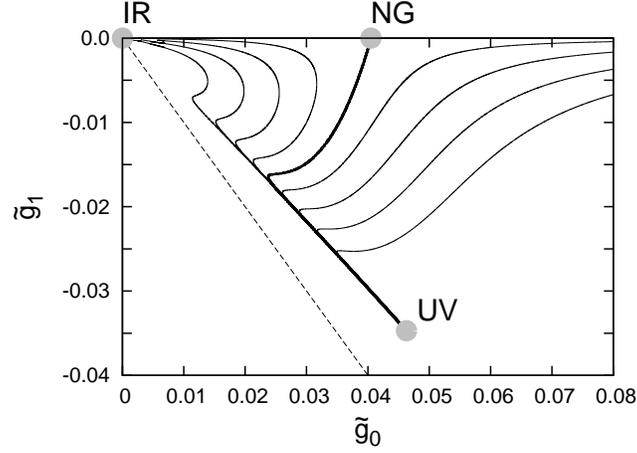,width=6cm,angle=-90}
\caption{\label{fig:nlsmphase} The theory space of the nonlinear $\sigma$ model is presented.
The trajectories start from the vicinity of the UV NGFP. The ones which tend to the left towards the
IR fixed point are in the broken phase, the others belong to the symmetric phase.
The fixed points are denoted by gray points, they
are: UV attractive NGFP $\to$ {\it UV}, hyperbolic NGFP $\to$ {\it NG}, and
IR fixed point $\to$ {\it IR}. The separatrix is situated between the UV and the NG
fixed points, and is denoted by the thick line. The latter is a saddle point and plays the role
of the crossover fixed point in this model. The dashed line shows the singularity limit,
when $\t g_0=-\t g_1$.}
\end{center}
\end{figure}

Taking into account more interacting terms in the action in \eqn{nlsm_action} we should
introduce further couplings in the model \cite{Flore:2012ma}. The $\beta$ functions for
two couplings reads as
\bea\label{nlsmbeta}
\t\beta_{\t g_0} &=& \t g_0+\t g_0(N-2)\t Q_{d/2,2}+d\t g_1(N-2)\t Q_{d/2+1,2},\nn
\t\beta_{\t g_1} &=& -\t g_1+\t g_1(N-2)\t Q_{d/2,2},
\eea
where
\beq
\t Q_{n,l} = \frac1{(4\pi)^{d/2}\Gamma(n)}\left(\frac{(2n+2+\partial_t)\t g_0}
{n(n+1)(\t g_0+\t g_1)^l}+\frac{2(2n+4+\partial_t)\t g_1}{n(n+2)(\t g_0+\t g_1)^l}\right).
\eeq
The phase structure is given in \fig{fig:nlsmphase}.
The model has two phases henceforward, but new fixed points appear. At $\t g_{0NG}=2/5\pi^2$ and
$\t g_{1NG}=0$ a hyperbolic NGFP appears. The trajectories are UV attracted from the direction
of $\t g_0$ and UV repelled from $\t g_1$. It implies that including a new coupling into the model
the asymptotic safety obtained for a single coupling disappears. The eigenvalues are
$s_{NG0}=-6/5$ and $s_{NG1}=2$. The former gives $\nu=-1/s_{NG0}=5/6$ for the critical
exponent, which coincides with the one obtained for $\t g_0$ alone. However a further NGFP
appears at $\t g^*_{UV0}=16/35\pi^2$ and $\t g^*_{UV1}=-12/35\pi^2$. The new fixed point is
UV attractive as the flows show in \fig{fig:nlsmphase}, the eigenvalues are $s_{UV0}=-0.457$
and $s_{UV1}=-13.11$. Let us mention, that a further hyperbolic fixed point appears in the
theory space far from the region shown in \fig{fig:nlsmphase}.
Noticeably the trajectories meet at the GFP in the broken phase, if one
follows the evolutions into the deep IR region, see \fig{fig:nlsmphase}.

\begin{figure}
\begin{center} 
\epsfig{file=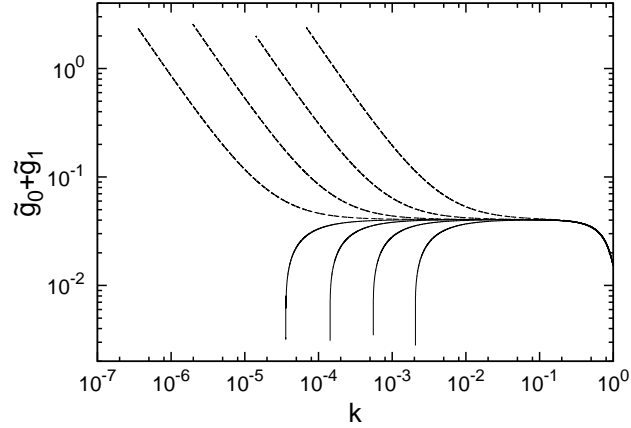,width=6cm,angle=-90}
\caption{\label{fig:nlsmsing} The flow of the quantity $\t g_0+\t g_1$ is shown.
The full (dashed) line corresponds to evolutions in the broken (symmetric) phase, respectively.
}
\end{center}
\end{figure}

According to
the $\beta$ functions in \eqn{nlsmbeta} the quantity $\t g_0+\t g_1=0$ makes the flows
singular. We plotted it for different UV initial values of the couplings in \fig{fig:nlsmsing}.
The trajectories have three different scaling regimes. In the UV there is a short, relevant
scaling region induced by the UV attractive NGFP. Then the scalings become marginal in the
vicinity of the hyperbolic NGFP. In the IR region the trajectories belonging to the symmetric
phase diverge as $k^1$.
In the broken phase the flows tends to zero sharply at a certain scale $k_c$, which can be
identified with the correlation length $\xi$ again. Therefore one can determine the exponent
$\nu$ of $\xi$ in the IR. It gives $\nu=0.835$, which is very close to the value $\nu=5/6$, that
was got analytically in the vicinity of the hyperbolic NGFP.

\subsection{The Gross-Neveu model}

The Gross-Neveu (GN) model is a 2-dimensional quantum field theoretic
model of $N_f$ flavors of massless relativistic fermions which interact via a four-fermionic
term \cite{Gross:1974jv}. The model is asymptotically free. The GN model was also
investigated at finite chemical potential in $d=2$, where an analytic solution of the crystal
ground state is found \cite{Thies:2003br}. In $d=3$ the model is not asymptotically free
\cite{Braun:2010tt}. It can also be studied at finite temperature \cite{Castorina:2003kq} and
chemical potential \cite{Hands:1995jq}.
The Euclidean effective action of the GN model has the form \cite{Braun:2010tt}
\beq
S[\bar\psi,\psi] = \int_x\left[Z_\psi\bar\psi i \sla{\partial}\psi+\frac{\bar g}{2N_f}
(\bar\psi \psi)^2\right],
\eeq
with the wavefunction renormalization $Z_\psi$ which is set to $Z_\psi=1$ in LPA.
The dimensionless four-fermion coupling $g$ can be obtained from the dimensionful $\bar g$ as
\beq
g=Z_\psi^{-2}k^{d-2}\bar g.
\eeq
The functional RG treatment of the GN model leads to the RG flow equation
\beq\label{gnbetag}
\beta_g=(d-2+2\eta_\psi)g-4d_\gamma v_d l_1^F(0) g^2
\eeq
in the $N_f\to\infty$ limit \cite{Braun:2010tt}. The constant $d_\gamma$ denotes the
dimension of the Dirac-algebra, e.g. it is $d_\gamma=4$ when $d=3$. The other constant $l_1^F(0)$
is an IR regulator dependent quantity, for example $l_1^F(0)=2/d$ when one uses
Litim's regulator. The model has two fixed points. We have a GFP at $g^*=0$ with the eigenvalue
$s_G=d-2$. It shows that the four-fermion coupling multiplies an irrelevant operator, thus
the model is perturbatively nonrenormalizable. The other NGFP is situated at
\beq\label{gnngfp}
g^*=\frac{d-2+2\eta_\psi}{4d_\gamma v_d l_1^F(0)},
\eeq
which, by using Litim's regulator in LPA and considering $d=3$ becomes $g^*=3\pi^2/4$.
Its corresponding eigenvalue is $s_{UV}=2-d$, which gives a relevant coupling when $d>2$, so
this fixed point is a UV attractive NGFP. In the case of $d=2$ the model is asymptotically free,
perturbatively renormalizable with its UV attractive GFP.

\begin{figure}
\begin{center} 
\epsfig{file=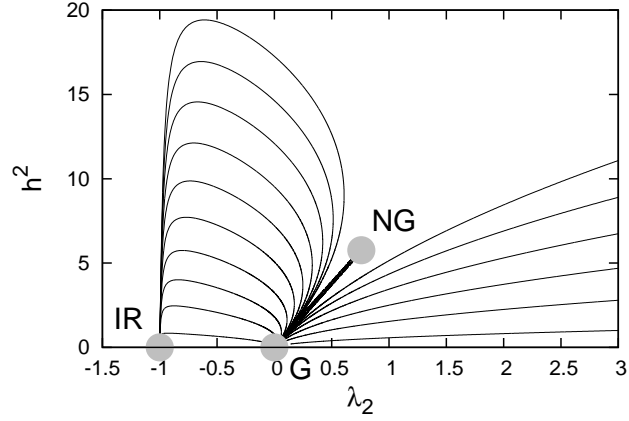,width=6cm,angle=-90}
\caption{\label{fig:gnphase} The theory space of the 3d Gross-Neveu model is shown.
The trajectories tending to the IR fixed point are in the broken phase, the others
belong to the symmetric phase. The fixed points are denoted by gray points. The
separatrix between the phases is represented by the thick line and connects the 
Gaussian (G) and the non-Gaussian (NG) fixed points. The GFP is a UV attractive, the hyperbolic
NGFP is a crossover fixed point.}
\end{center}
\end{figure}

\begin{center}
\begin{figure}[ht]
\epsfig{file=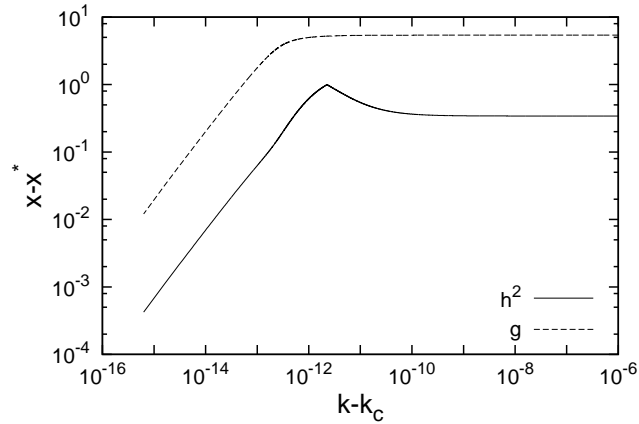,width=6cm,angle=-90}
\caption{\label{fig:gnh2}
The scaling of the couplings in the 3d Gross-Neveu model is shown, $x$ denotes $h_2$ and
$\lambda_2$, the fixed points values are $h_2^*=0$ and $\lambda_2^*=-1$. The couplings
have a long marginal scaling region induced by the hyperbolic NGFP. In the IR limit there
is another scaling regime, which can be uncovered by using the shifted scale $k-k_c$.}
\end{figure}
\end{center}

In the partially bosonized version of the GN model
\cite{Schon:2000he,Castorina:2003kq,Braun:2010tt}  a hyperbolic fixed point, and a UV
attractive GFP appears, so the model becomes asymptotically free.
From \cite{Braun:2010tt} the corresponding evolution equations for the d-dimensional
bosonized GN model is
\bea
\dot u &=& -du+(d-2+\eta_\sigma)u'\rho-2d_\gamma v_d l_0^{(F)d}(2h^2\rho;\eta_\psi)
+\frac1{N_f}2v_dl_0^d(u'+2\rho u'';\eta_\sigma),
\eea
where we impose a polynomial ansatz for the potential
\beq
u(\rho) = \sum_{n=0}^\infty \frac{\lambda_{2n}}{n!}\rho^n,
\eeq
with the bosonic couplings $\lambda_{2n}$. One can relate the coupling $g$ in \eqn{gnbetag}
as $g=h^2/\lambda_2$. The flow equation for the Yukawa coupling is
\beq
\dot h^2 = (d-4+2\eta_\psi+\eta_\sigma)h^2+\frac1{N_f}8v_dh^4
l_{1,1}^{(FB)d}(0,\lambda_2;\eta_\psi,\eta_\sigma).
\eeq
The anomalous dimensions are
\bea
\eta_\sigma &=& 8\frac{d_\gamma v_d}{d} h^2 m_4^{(F)d}(0;\eta_\psi),\nn
\eta_\psi &=& \frac1{N_f}8\frac{v_d}{d} h^2 m_{1,2}^{(FB)d}(0,\lambda_2;\eta_\psi,\eta_\sigma).
\eea
We introduced the threshold functions
\bea
l_0^d(\omega;\eta_\sigma) &=& \frac2{d}\left(1-\frac{\eta_\sigma}{d+2}\right)\frac1{1+\omega},\nn
l_0^{(F)d}(\omega;\eta_\psi) &=& \frac2{d}\left(1-\frac{\eta_\psi}{d+2}\right)
\frac1{1+\omega},\nn
l_{1,1}^{(FB)d}(\omega_1,\omega_2;\eta_\psi,\eta_\sigma) &=&
\frac2{d}\frac1{(1+\omega_1)(1+\omega_2)}\left\{\left(1-\frac{\eta_\psi}{d+1}\right)
\frac1{1+\omega_1}+\left(1-\frac{\eta_\sigma}{d+2}\right)\frac1{1+\omega_2}\right\},\nn
m_4^{(F)d}(\omega;\eta_\psi) &=& \frac1{(1+\omega)^4}+\frac{1-\eta_\psi}{d-2}\frac1{(1+\omega)^3}
-\left(\frac{1-\eta_\psi}{2d-4}+\frac14\right)\frac1{(1+\omega)^2},\nn
m_{1,2}^{(FB)d}(\omega_1,\omega_2;\eta_\psi,\eta_\sigma) &=&
\left(1-\frac{\eta_\sigma}{d+1}\right)\frac1{(1+\omega_1)(1+\omega_2)^2}.
\eea
Keeping only the couplings $h^2$ and $\lambda_2$ we get the evolution equations for $d=3$
\bea
\dot \lambda_2 &=& -2\lambda_2+\frac4{3\pi^2}h^2+\frac5{3\pi^2}h^2 \lambda_2,\nn
\dot h^2 &=& -h^2+\frac5{3\pi^2}h^4+\frac{2h^4(2+\lambda_2)
-\frac2{9\pi^2}h^6}{N_f 3\pi^2(1+\lambda_2)^2}.
\eea
The flow equations have a UV attractive GFP at the origin at $h^{2*}_G=0$ and
$\lambda_{2G}^*=0$ and a
non-Gaussian saddle point at $h^{2*}_{NG}=5.764$ and $\lambda_{2NG}^*=0.758$ for $N_f=12$.
The phase structure can be seen in \fig{fig:gnphase}. The trajectories tending to
left and right correspond to regions of different phases in the model.
After the reparameterization of the couplings according to $\omega=1+\lambda_2$, $\chi=h^2/\omega$
and $\partial_\tau=\omega \partial_t$ we get
\bea
\partial_\tau \omega &=& 2\omega(1-\omega)+\frac{\chi\omega^2}{3\pi^2}(5\omega-1),\nn
\partial_\tau \chi &=& 2\chi(\omega-2)+\frac{\chi^2}{18\pi^2}\left(7\omega+1
-\frac{\chi\omega}{3\pi^2}\right).
\eea
The new flow equations have three physical fixed points. The UV attractive GFP can be
found at $\omega^*_G=1$ and $\chi^*_G=0$. The NGFP is situated at $\omega^*_{NG}=1.758$ and
$\chi^*_{NG}=3.278$ with a positive and a negative eigenvalues. There is a third fixed point at
$\omega^*_{IR}=0$ and $\chi^*_{IR}=355.206$ with two positive eigenvalues implying that
that this fixed point is a UV repulsive or IR attractive, therefore we can identify it
with the IR fixed point at $h^{2*}_{IR}=0$ and $\lambda^*_{2IR}=-1$ if we express them
in terms of the original couplings. One can investigate the scaling of the couplings in the
vicinity of the IR fixed point, and it is shown in \fig{fig:gnh2}.
The couplings are constants in the beginning of the flows, giving marginal scaling due to the
hyperbolic NGFP in \fig{fig:gnphase}, that is $x-x^*\sim (k-k_c)^0$. In the deep IR region the
couplings have singular behaviors, i.e. they abruptly become zero or infinite at a certain scale
$k_c$. If one plots the couplings as the function of the shifted scale $k-k_c$, then one can
obtain a power-law like behavior according to $x-x^*\sim (k-k_c)^1$, where $x^*$ is the fixed
point value of the couplings, for example $h_2^*=0$ and $\lambda_2^*=-1$, see \fig{fig:gnh2}.
It implies that the original four-fermion coupling $g$ tends to zero in the IR limit.
We also calculated the critical exponent $\nu$ of the correlation length in the IR regime by
the previously shown method, which is based on identifying the correlation length by the
reciprocal of the stopping scale $k_c$. The results are summarized in \tab{tab:gnnu}. They show
good coincidence with the ones obtained in
\cite{Braun:2011pp}.
\begin{table}
\begin{center}
\begin{tabular}{|c||c|c|c|c|c|c|}
\hline
  $N$ & 1 & 2 & 6 & 12 & 50 \\
\hline
 $\nu_{\mbox{IR}}$ & 0.922 & 0.976 & 0.990 & 0.996 & 1.00 \\
\hline
\end{tabular}
\end{center}
\caption{\label{tab:gnnu} The critical exponent $\nu$ in the 3d Gross-Neveu model
for various values of $N$.}
\end{table}

\subsection{The 2d sine-Gordon model}

The 2d sine-Gordon model can be an example for asymptotic freedom and asymptotic safety,
simultaneously. Its effective action contains a sinusoidal potential of the form
\beq
\Gamma_k = \int\left[\frac{z}2 (\partial_\mu\phi)^2{+}u \cos\phi\right],
\eeq
where $z$ is the field independent wavefunction renormalization and $u$ is the coupling.
The perturbative RG results beyond LPA \cite{Amit:1979ab} can account for the KT phase
transition and give $1/z\to 0$ for $1/z<8\pi$ in the IR limit. By using the flow 
equation approach \cite{Kehrein:1999nx,Kehrein:2000ap} a different IR limit is obtained for
the frequency, i.e. $1/z\to 4\pi$. However we note, that this method
cannot recover the leading order  perturbative UV results for $1/z<4\pi$,
due to the opposite sign obtained for the evolution equation for 
the frequency. Functional RG approaches can map the phase structure of the SG model
in LPA \cite{Nandori:1999vi,Nandori:2003pk,Nagy:2006pq,Pangon:2009wk,Nandori:2009ad,
Nandori:2010xr} and can also take into account the evolution of the wavefunction renormalization
\cite{Nandori:2000rx,Nandori:2002ri,Nagy:2009pj,Nagy:2010mf,Nandori:2011ss}.

Besides the $Z_2$ symmetry the model has a periodic symmetry. The model has two phases.
The effective potential should be convex (nonconcave), furthermore the RG equations should not
break the periodic symmetry, too. This two requirements can be satisfied if the effective
potential is flat, or zero \cite{Nandori:1999vi,Alexandre:2010gh}.
Then how one can distinguish the phases in the model?
The simple answer is that one should consider the dimensionless effective potential,
since only the dimensionful one should be convex. In the symmetric phase the dimensionless
effective potential is flat, while it is a concave function in the broken phase. In LPA
we got that it is $\t V_0 = -\phi^2/2$, which is repeated periodically in the field variable.
This result can be obtained if one considers the evolution of higher harmonics in the potential,
or if one follows the flow of the potential without its Fourier expansion.

The phases cannot be always distinguished by the form of the effective potential. The real
difference comes from the property how the effective potential depends on the UV
initial value of the couplings. It seems to contradict to the fact
that the effective potential is a simple function at $k=0$. However this scale
cannot be reached, since during the evolution we use that the ratio $\Delta k/k$
is small disabling us to reach the exact value of $k=0$. If we consider the RG time $t$
then it is more apparent that we have only tendencies to infinity. Therefore it is
meaningful to investigate how the asymptotic behavior of the effective action becomes
the effective potential when the scale $k$ is lowered. One can observe that
a power law scaling appears in the deep region IR which can imply that the appearing scaling
behavior can be extrapolated to $k=0$.
In the case of the 2d SG model the effective potential in the symmetric phase
depends on the initial value of $\t u(k_\Lambda)$,
while it is universal (independent on the initial coupling) in the broken phase.
This type of separation of phases always works in any models.

\subsubsection{Local potential approximation}

By using the power law IR regulator in \eqn{regpow} with the choice $b=1$
the leading order approximation of the RG evolution equations is
\bea
\label{sgtree}
\dot{\t u} &=& -2 \t u,\nn
\dot z &=& 0.
\eea
These equation are usually referred to as UV scalings.
The fixed points are $\t u^*=0$ with arbitrary $z^*$. These points constitute
a line of fixed points. Since $z$ does not evolve, then we are in LPA.
The matrix $M$ corresponding to the linearized RG equations in the vicinity of the
line of fixed points is
\beq
\begin{pmatrix}
-2+\frac1{4\pi z} & -\frac{\t u}{4\pi z^2}\\
0 & 0\\
\end{pmatrix}_{\t u^*=0,z^*}
=
\begin{pmatrix}
-2+\frac1{4\pi z^*} & 0\\
0 & 0\\
\end{pmatrix}
\eeq
The eigenvalue to the evolution of $\t u$ is $s=-2+\frac1{4\pi z^*}$.
Its sign depends on the value of $z^*$. If $z^*<1/8\pi$ then $s>0$ and the
coupling is irrelevant.  When $z^*>1/8\pi$ then $s<0$ and the
coupling scales in relevant manner. These two types of scalings give the two phases
of the model separated by the critical value $z^*_c=1/8\pi$. The point $\t u^*_c=0$ and
$z^*_c=1/8\pi$ is called the Coleman point. The separatrix is a vertical line
and goes through the Coleman point in the theory space.

The symmetric phase, when $z^*<1/8\pi$ the phase is usually called the nonrenormalizable
(or massless) phase due to the irrelevant scaling of the coupling $\t u$. The other
phase is the broken one. There the $Z_2$ symmetry breaks down, it is called the renormalizable
(or massive) phase ($\t u$ is relevant there).

\subsubsection{Linearized RG equations}

\begin{figure}
\begin{center}
\includegraphics[width=4.5cm,angle=-90]{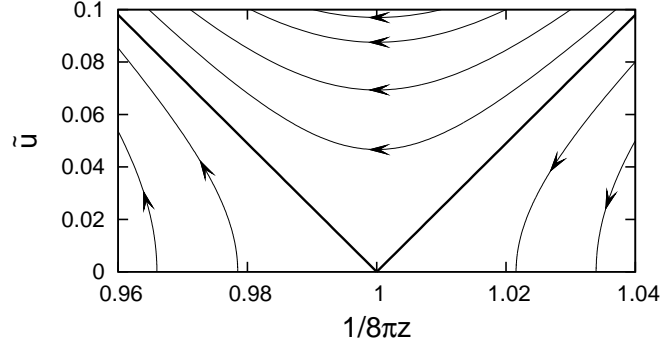}
\caption{\label{fig:sgpt}
The phase structure of the 2d sine-Gordon model treated in linear approximation. The separatrix
is denoted by a thick line.
} 
\end{center}
\end{figure}

By using the power law IR regulator in \eqn{regpow} with the choice $b=1$
the linearized RG evolution equations are
\bea
\label{sgpt}
\dot{\t u} &=& \t u\left(-2+\frac1{4\pi z}\right)+\ord{\tilde u^2},\nn
\dot z &=& -\frac{\tilde u^2}{24\pi}+\ord{\tilde u^3}.
\eea
The fixed point solution is $\t u^*=0$ and $z^*$ arbitrary. The matrix $M$ is
\beq
\begin{pmatrix}
-2+\frac1{4\pi z} & -\frac{\t u}{4\pi z^2}\\
-\frac{\tilde u}{12\pi} & 0\\
\end{pmatrix}_{\t u^*=0,z^*}
=
\begin{pmatrix}
-2+\frac1{4\pi z^*} & 0\\
0 & 0\\
\end{pmatrix}.
\eeq
In case of the linear approximation the wavefunction renormalization $z$ does not evolve,
similarly to the LPA case, and we also have the same eigenvalues.
It also implies that in the vicinity of the line of fixed points $z$ does not evolve, which is
reflected by perpendicular trajectories to the horizontal axis.
The eigenvalues are the same as was got in LPA. The theory space is presented in \fig{fig:sgpt}.
The point $z^*_c=1/8\pi$ and $\t u^*_c=0$ is now called the Kosterlitz-Thouless (KT) fixed point.
There is a separatrix which divides the theory space into two parts. The trajectories tending to
the line of fixed points belong to the symmetric phase, all other trajectories constitute the
broken phase. Around the KT point an essential scaling appears due to the infinite order phase
transition in the model.

\subsubsection{Exact RG equations}

\begin{figure}
\begin{center}
\includegraphics[width=4.5cm,angle=-90]{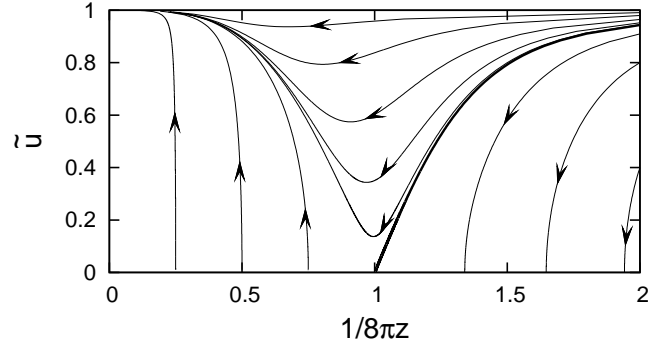}
\caption{\label{fig:sgrg}
The phase structure of the 2d sine-Gordon model. The separatrix is denoted by a thick line.
} 
\end{center}
\end{figure}

The RG equations can be obtained exactly when $b=1$ in this approximation (where the
potential contains only the fundamental mode, and the wavefunction renormalization is
field independent). The flow equations are
\bea
\label{sgrg}
\dot{\t u} &=& -2\t u+\frac1{2\pi\tilde u z} \left[1-\sqrt{1-\tilde u^2}\right],\nn
\dot z &=& -\frac1{24\pi}\frac{\tilde u^2}{(1-\tilde u^2)^{3/2}}.
\eea
The phase structure is plotted in \fig{fig:sgrg}.
One can easily see that the phase structure of the linearized treatment in \fig{fig:sgpt}
is a part of this theory space around the KT point. The separatrix was a straight line there
which is now curved. If one tries to find the fixed points of the model then immediately
realizes that it has not got any. However the phase structure suggests, that there is the
line of fixed points at $\t u^* = 0$, and it seems that we have an IR fixed point
in the broken phase at $\t u^*=1$ and $1/z^*=0$. Furthermore a UV attractive NGFP appears
at $\t u^*=1$ and $z^*=0$. Let us see them one by one.

\begin{center}
{\it Line of fixed points}
\end{center}
The line of fixed points can be found at $\t u^* = 0$ with arbitrary $z^*$. Its scaling
behavior can be obtained by expanding the RG flow equations in \eqn{sgrg} in $\t u$ which
gives back the approximate equations in \eqn{sgpt}. Thus we have the same scaling behavior, i.e.
a relevant scaling for $z^*>1/8\pi$ and an irrelevant one for $z^*<1/8\pi$ and the different
scaling regimes are separated by the KT point at $\t u^*_c = 0$ and $z^*_c=1/8\pi$.

\begin{center}
{\it IR fixed point}
\end{center}

The IR fixed point is situated at $\t u^*=1$ and $1/z^*=0$ in the broken phase. It cannot be
found directly from equations in \eqn{sgrg}, too, therefore we should reparameterize the
couplings. After introducing $\omega = \sqrt{1-\t u^2}$, $\chi=1/z\omega$
and $\partial_\tau = \omega^2 k\partial_k$ we arrive at the evolution equations
\bea
\partial_\tau\omega &=& 2\omega(1-\omega^2)-\frac{\omega^2\chi}{2\pi}(1-\omega),\nn
\partial_\tau\chi &=& \chi^2\frac{1-\omega^2}{24 \pi}
-2\chi(1-\omega^2)+\frac{\omega\chi^2}{2\pi}(1-\omega),
\eea
possessing the fixed point $\omega^*=0$ and arbitrary $\chi^*$ which corresponds
to $\t u^*=0$ and arbitrary $z^*$, thus we get back the lines of fixed points. However
there is another fixed point at $\chi^*=0$ and $\omega^*=0$, which can be identified
by the IR fixed point at $1/z^*=0$ and $\t u^*=1$, which is IR attractive.

If we introduce $\bar k =\min(z p^2+R)$, the RG evolution becomes singular at $k=k_c$ when
\beq\label{degen}
\left.\bar k^2 - V''_k(\phi=0)\right|_{k=k_c}=0,
\eeq
where $\bar k^2 = b k^2[z/(b-1)]^{1-1/b}$, when $b=1$, then $\bar k=k$. The solution of this
equation defines the scale where the potential becomes degenerate.

The normalized coupling $\bar u$ tends to 1 for every value of $b$. It shows that the degenerate
potential (which satisfies \eqn{degen}) occurs in the IR limit of the broken
phase independently of the RG scheme. In the symmetric phase the evolution of $z$ is negligible
in the IR giving the same evolution as was obtained in LPA with the line of fixed points.

\begin{figure}[ht]
\begin{center}
\epsfig{file=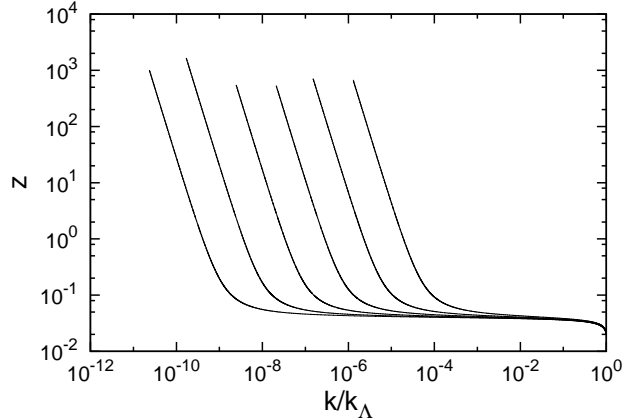,width=6cm,angle=-90}
\caption{\label{fig:sgzk}
The scaling of the wavefunction renormalization $z$ is plotted as the function of $k$.
The initial values are: $z(k_\Lambda)=0.6$ and $\t u(k_\Lambda)$ around 1.95.
}
\end{center}
\end{figure}

\begin{figure}[ht]
\begin{center}
\epsfig{file=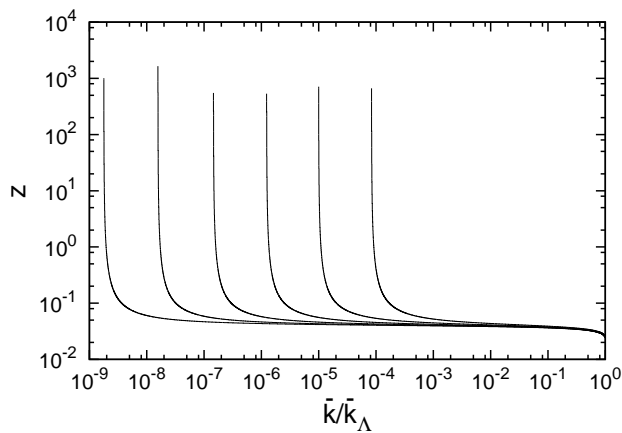,width=6cm,angle=-90}
\caption{\label{fig:sgzkbar}
The scaling of the wavefunction renormalization $z$ is plotted as the function of $\bar k$.
The initial values are: $z(k_\Lambda)=0.6$ and $\t u(k_\Lambda)$ around 1.95.
}
\end{center}
\end{figure}

\begin{figure}[ht]
\begin{center}
\epsfig{file=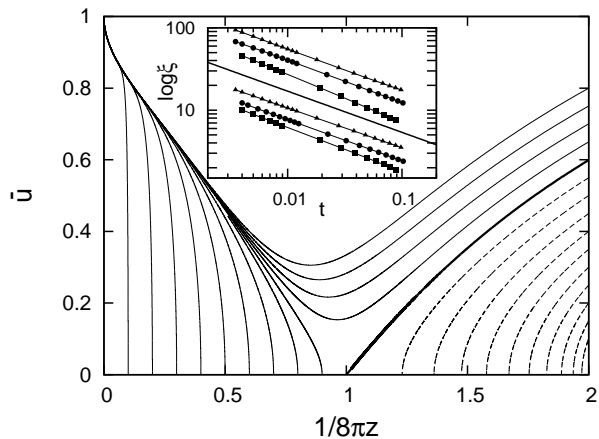,width=6cm,angle=-90}
\caption{\label{fig:sgphb5}
Phase diagram of the SG model, with $b=5$. The dashed (solid) lines represent
the trajectories belonging to the (broken) symmetric phase, respectively.
The wide line denotes the separatrix between the phases. The KT point can be considered as
the crossover fixed point. The inset shows the
scaling of the correlation length as the function of the reduced temperature $t$. The curves
are shifted for better visibility. The lower (upper) set of lines corresponds to the IR (KT)
fixed point. The triangle, circle and square correspond to $b=2,5,10$, respectively.
In the middle a straight line with the slope $-1/2$ is drawn to guide the eye.
}
\end{center}
\end{figure}

One can easily show that the critical exponent $\eta_v$ characterizing the vortex--vortex
correlation function \cite{Kosterlitz:1974sm} is $\eta_v=1/4$ independently of the parameter $b$
\cite{Nagy:2009pj}. However the anomalous dimension for the
correlation function of the field variables gives $\eta=0$ in the vicinity of the KT point.
In the deep IR scaling region the situation changes significantly, there new scaling laws appear.
\fig{fig:sgzk} shows that around the KT point (at about $k/\lambda\sim 10^{-4}$) $z$ 
is practically constant, giving $\eta=0$, while in the IR region $z$ starts to diverge
according to a power law scaling.
This scaling regime is induced by the IR fixed point.
The corresponding exponent depends on the scheme, i.e. on the value of $b$. The coupling
$\bar u$ shows the same qualitative behavior. If one plots $z$ as the function of $\bar k$
then one obtains the results of \fig{fig:sgzkbar}.
The flows show sharp singularities as the function of the scale $\bar k$ so as the
other couplings, making the evolution scheme independent. The modes which are integrated
out during the RG procedure can be indexed not only by the scale $k$ but the scale $\bar k$ and
one has a smallest value of the scale $\bar k_c$. Again one can identify the correlation length
according to $\xi=1/\bar k_c$, furthermore the critical exponent also can be identified in a
similar manner as was introduced in the d-dimensional $O(N)$ model.
In the 2d SG model the temperature is proportional to $z_{k_\Lambda}$, and
thus the reduced temperature $t$ is $t\sim z_{k_\Lambda}-z_{c~k_\Lambda}$, where
$z_{c~k_\Lambda}$ is the initial value of the wavefunction renormalization at the separatrix.
In order to demonstrate whether the phase structure is qualitatively
the same we plotted it for $b=5$ in \fig{fig:sgphb5}.
In its inset we plotted $\log \xi$ as the function of $t$. We obtained a straight line
implying the scaling of $\xi$ according to
\beq\label{nukt}
\log\xi \propto t^{-\nu}.
\eeq
There are two types of correlation lengths, one is defined as usual, i.e. around the KT
turning point of the coupling $\bar u$. Another one is identified as $\xi=1/\bar k_c$ in
the neighborhood of the IR fixed point.  It can be seen from the inset of \fig{fig:sgphb5}
that the scaling of $\xi$ shows an infinite order phase transition for all schemes with the
exponent $\nu\approx 0.5$.

We obtained that the exponent $\nu$ coincides calculated around the KT and IR points.
This is due to fact that condensate is a global object in the broken phase
which is perceptible at practically any scale. The other exponents do not necessarily
coincide at the KT and the IR points, e.g. the anomalous dimension $\eta=0$ around
the KT point, while it is $\eta=2b/(b-1)$ around the IR one.

The singularity at low scales appears because the condition in \eqn{degen} is satisfied.
The condition shows that there are many modes in the model with infinitesimally small, practically
zero energies, they are called soft modes.
The ground state of the model can pick up arbitrary number of these fluctuations,
which makes the ground state degenerate. The huge amount of soft modes constitute the
condensate in the broken phase. They signal that the microscopic degrees of freedom are
not suitable to describe the model anymore in the broken phase at those small energy scales,
one should turn to new field variables. It is reminiscent of the composition of hadrons
from quarks, which are the original degrees of freedom in QCD.

One can conclude that the IR fixed point provides us the low energy limit of the theory under
investigation. On the other hand the soft modes suggests that the quantum-classical transition
appears in the broken phase, however correct treatment should be discussed by the closed time path
method \cite{Polonyi:2011yx,Polonyi:2012tj}.

\begin{center}
{\it UV NGFP}
\end{center}

The UV NGFP also cannot be seen directly analytically but it can be easily recognized in the
top right corner of the phase diagram in \fig{fig:sgrg}. The LPA approximation trivially shows
that in the UV attractive NGFP the model has a singularity, as in the IR fixed point. Furthermore,
from the RG equations in \eqn{sgrg} the ratio $du/dz$ can be calculated, and it is easy
to show that the ratio is zero if $z\to 0$ and $\t u\to 1$. This point of the phase
space corresponds to the UV NGFP of the 2d SG model and it also makes the RG equations
singular. The UV singularity signals the upper limit of the applicability of the model. It implies that at
high energies the model requires new elementary excitations. One can make it more apparent
in the framework of the XY model, which is in the same universality class as the 2d SG model.
There the excitations are represented by vortices consisting of concentric forms of spins
\cite{Huang:1990via}. The blocking towards the higher scales means that the vortices
have smaller and smaller diameter. Naturally at a certain scale we reach a limit, where
the vortex reduces to a single spin. At this scale the new elementary excitations should
be the single spins instead of the original vortices. Therefore in the UV limit the system of the
charged vortices should be replaced by a neutral spin system.

The 2d SG model shows a nice example, where the functional RG method gives us both the high
and low energy scale limits of its applicability. The low energy limit is usually indicated
by the IR fixed point as was demonstrated in the article for many models. There the
global interaction belonging to the condensate of the broken phase becomes IR relevant
and introduces new excitations. However there should be a UV limit of the models, since
it seems nonphysical to assume that a model can be valid at arbitrarily high energies.
So far the functional RG method gave the upper UV limit only for the 2d SG model and its
generalizations (massive \cite{Nagy:2006ue,Nandori:2010ij} or layered
\cite{Nandori:2005xa,Nandori:2005pa,Nandori:2005fj,Nandori:2006tv,Nandori:2007zs,Nagy:2012qz}
SG models).

\subsection{Quantum Einstein gravity}

The QEG is the quantum field theoretic model of gravity \cite{Reuter:2002kd,Reuter:2012id}.
There the metric fields are considered as fundamental degrees of freedom, and
play the role of the field variable, thus the model can be treated in the
framework of path integral formalism. By using the Einstein-Hilbert truncation the model
contains the Newton constant $G_k$ and the cosmological constant $\Lambda_k$ as couplings.
According to their canonical dimensions the dimensionless cosmological constant
$\lambda = k^{-2}\Lambda_k$ is relevant, i.e. perturbatively renormalizable, however the
dimensionless Newton constant $g$ scales in irrelevant manner, since $g = G k^{d-2}$. Thus the
model seems to be non-renormalizable according to perturbative considerations. This might be the
signal of the necessity to introduce new dynamical variables for elementary excitations, e.g.
strings.

However the concept of asymptotic safety may give a helping hand to QEG.
It was shown that the model seems to possess a UV attractive NGFP, which makes QEG renormalizable
\cite{Weinberg:1979,Reuter:1996cp,Reuter:2007rv,Lauscher:2001rz,Reuter:2001ag}
and shows asymptotic safety \cite{Lauscher:2001cq,Niedermaier:2006ns,Percacci:2007sz,
Percacci:2011fr,Nink:2012vd,Bonanno:2012jy,Harst:2012ni}.
It is a great challenge to find experimental evidence for quantum gravity, however
there are promising ideas to catch its effects in LHC \cite{Dobrich:2012nv,Eichhorn:2012kk}.
Usually the model is considered in $d=4$, however the 3-dimensional \cite{Demmel:2012ub}
and higher dimensional cases \cite{Litim:2003vp,Fischer:2006fz,Litim:2008tt} are also
investigated. In the context of dimension we notice that a dimensional reduction
takes place during the flow to small distances from $4\to 2$
\cite{Reuter:2011ah,Calcagni:2012qn,Reuter:2012xf,Reuter:2012id}.
Note that the dimensional reduction refers to the spectral dimension only;
the topological dimension does not undergo a dimensional reduction.
The bare action of QEG belongs to the UV attractive NGFP of the RG flows
\cite{Reuter:2012id,Reuter:2012xf}. Some kind of classical limit can be associated to any
fixed points, which can give different General Relativity and cosmology \cite{Bonanno:2012jy}.

QEG is defined through a diffeomorphism invariant functional of the metric $g_{\mu\nu}$ in QEG,
and the background field technique is used to preserve the gauge symmetry when the RG flow
equations are derived. Here we flash the main points of the derivation given in
\cite{Reuter:1996cp}. The functional integral should be performed over all metrics
$\gamma_{\mu\nu}$. It is decomposed into
\beq
\gamma_{\mu\nu}(x)=\bar g_{\mu\nu}(x)+h_{\mu\nu}(x)
\eeq
where $\bar g_{\mu\nu}(x)$ is the background field metric and $h_{\mu\nu}(x)$ is the new
integration variable. The generating functional is
\beq\label{qegblact}
\exp\{W_k\} = \int{\cal D}[h_{\mu\nu};{\cal C};\bar{\cal C}]
\exp\{-S[\bar g+h]-S_{gf}[h;\bar g]-S_{gh}[h,{\cal C},\bar {\cal C};\bar g]
-\Delta_k S[h,{\cal C},\bar {\cal C};\bar g]-S_{source}\},
\eeq
where $S[\bar g+h]=S[\gamma]$
is the classical action, which is invariant under the general coordinate transformation
(${\cal L}_v$ is the Lie derivative w.r.t. the vector field $v$)
\beq
\delta\gamma_{\mu\nu}={\cal L}_v\equiv v^\rho\partial_\rho\gamma_{\mu\nu}+
\partial_\mu v^\rho\gamma_{\rho\nu}+\partial_\nu v^\rho\gamma_{\mu\rho}.
\eeq
The term $S_{gf}$ in \eqn{qegblact} denotes the gauge fixing term
\beq
S_{gf}[h;\bar g] = \frac1{2\alpha}\int d^d x\sqrt{\bar g}\bar g^{\mu\nu} F_\mu F_\nu,
\eeq
where $F_\mu$ is linear in the field variable $h_{\mu\nu}$ and contains a certain
form of a first order differential operator of $\bar g_{\mu\nu}$ \cite{Reuter:1996cp}.
The term $S_{gh}$ stands for the Faddeev-Popov term with the ghosts $C^\mu$ and $\bar C_\mu$
\beq
S_{gh}[h,C,\bar C;\bar g]=-\frac1{\kappa}\int d^d x\bar C_\mu\bar g^{\mu\nu}
\frac{\partial F_\nu}{\partial h_{\alpha\beta}}{\cal L}_C(\bar g_{\alpha\beta}+h_{\alpha\beta}),
\eeq
with $\kappa=(32\pi G)^{-1/2}$. It is obtained by applying the gauge transformation
\beq
\delta h_{\mu\nu} = {\cal L}_v \gamma_{\mu\nu},~~\delta \bar g_{\mu\nu} = 0
\eeq
to $F_\mu$ and then replacing the parameters $v^\mu$ by the ghost field $C^\mu$.
The Faddeev-Popov determinant is represented by the path integrals over $C^\mu$ and $\bar C_\mu$.
The IR regulator should be applied to both the ghost and the gravitational field, and it reads as
\beq
\Delta_k S[h,C,\bar C;\bar g] = \hf\kappa^2\int d^d x\sqrt{\bar g}h_{\mu\nu}
{\cal R}^{grav}[\bar g]^{\mu\nu\rho\sigma}h_{\rho\sigma}
+\sqrt{2}\int d^d x\sqrt{\bar g}\bar C_\mu{\cal R}^{gh}[\bar g]C^\mu.
\eeq
The source terms in \eqn{qegblact} has the form
\beq
S_{source} = -\int d^d x\sqrt{\bar g}(t^{\mu\nu}h_{\mu\nu}+\bar\sigma_\mu C^\mu
+\sigma^\mu\bar C_\mu+\beta^{\mu\nu}{\cal L}_C(\bar g_{\mu\nu}+h_{\mu\nu})
+\tau_\mu C^\nu\partial_\nu C^\mu).
\eeq
In order to get the effective action one should introduce the classical field variables
\beq
\bar h_{\mu\nu}=\frac1{\sqrt{\bar g}}\frac{\delta W_k}{\delta t^{\mu\nu}},~~
\xi_\mu=\frac1{\sqrt{\bar g}}\frac{\delta W_k}{\delta \bar\sigma^\mu},~~
\bar \xi_\mu=\frac1{\sqrt{\bar g}}\frac{\delta W_k}{\delta \sigma^\mu},
\eeq
and then it can be obtained by the following Legendre transformation
\beq
\Gamma_k = \int d^d x \sqrt{\bar g}(t^{\mu\nu}\bar h_{\mu\nu}+\bar\sigma_\mu\xi^\mu
+\sigma^\mu\bar\xi_\mu)-W_k+\Delta S_k.
\eeq
Finally one arrives at the flow equation of the form
\beq
\dot\Gamma_k[\bar h,\xi,\bar \xi,\bar g]
=\hf\mbox{Tr}\frac{\sqrt{2}(\dot {\cal R}_k^{grav})_{\bar h\bar h}}
{(\sqrt{2}{\cal R}_k^{grav}+\Gamma_k'')_{\bar h\bar h}}
-\hf\mbox{Tr}\left[\kappa^2(\dot {\cal R}_k^{gh})_{\bar\xi \xi}
\left(\frac1{(\kappa^2{\cal R}_k^{gh}+\Gamma_k'')_{\bar\xi \xi}}
-\frac1{(\kappa^2{\cal R}_k^{gh}+\Gamma_k'')_{\xi\bar\xi}}\right)\right].
\eeq
A general form of the QEG effective action is
\beq\label{eaqeg}
\Gamma_k = \int d^d x\sqrt{\mbox{det}g_{\mu\nu}}
\left( \frac1{16\pi G_k}(2\Lambda_k-R)-\frac{\omega_k}{3\sigma_k}R^2+\ldots
+ \frac1{2\sigma_k} C^2+\frac{\theta_k}{\sigma_k}E+\ldots
+ \mbox{gf.~terms}+\mbox{gh.~terms}\right).
\eeq
The first terms come from the Taylor expansion of the general non-local functional $f(R)$
\cite{Machado:2007ea,Dietz:2012ic}. The first two terms constitute the Einstein-Hilbert
truncation, with the two dimensionless couplings $\lambda=\Lambda_k k^{-2}$ and $g=G_k k^{d-2}$.
Then the contributions coming from the square of the Weyl tensor
$C^2=C_{\mu\nu\rho\sigma}C^{\mu\nu\rho\sigma}$, and the Gauss-Bonnet topological invariant
$E=R_{\mu\nu\rho\sigma}R^{\mu\nu\rho\sigma}-4R_{\mu\nu}R^{\mu\nu}+R^2$
are considered \cite{Benedetti:2009iq} in \eqn{eaqeg}. In the end the gauge fixing
and the ghost terms are indicated. The functional $f(R)$ can have any form, e.g.
it can be a logarithmic or polynomial function \cite{Machado:2007ea}. The explicit
form of the evolution equations including the interaction term $R^2$ can be found in
\cite{Rechenberger:2012pm}. A further possible extension of QEG contains terms in the action in
\eqn{eaqeg} which describes interaction between the matter and the metric fields.
The matter field can be a fermionic system \cite{Eichhorn:2011pc,Eichhorn:2011ec,Dona:2012am}
a scalar field \cite{Percacci:2002ie,Percacci:2003jz,Zanusso:2009bs,Manrique:2010mq,
Manrique:2010am,Vacca:2010mj,Eichhorn:2012va}, or a gauge field
\cite{Folkerts:2011jz,Eichhorn:2011gc,Donkin:2012ud}. The improvement of the ghost sector is
also intensively studied \cite{Groh:2010ta,Eichhorn:2010tb}. 

From the extensions of the QEG effective action in \eqn{eaqeg} one can derive the
RG evolution equations for the couplings. Considering the 2-dimensional projection
of the theory space to the usual Newton and cosmological couplings we can obtain
different types of phase structures. However it has several universal properties, namely
that any extension of QEG shows two phases and possesses a UV attractive NGFP,
implying that model is asymptotically safe.

\subsection{Evolution equation}

Using Litim's regulator and the Einstein-Hilbert truncation, where the dimensionless
Newton and the cosmological couplings are considered, we obtain the following RG equations
\cite{Litim:2008tt}
\bea\label{gleveq}
\dot\lambda &=& -2\lambda+\frac{g}2 d(d+2)(d+5)
-d(d+2)\frac{g}2\frac{(d-1)g+\frac{1-4\lambda(1-1/d)}{d-2}}{g-g_b},\nn
\dot g &=& (d-2)g+\frac{(d+2)g^2}{g-g_b},
\eea
with
\beq
g_b = \frac{(1-2\lambda)^2}{2(d-2)}.
\eeq
One can introduce the gravitation anomalous dimension
\beq\label{etaqeg}
\eta = \frac{(d+2)g}{g-g_b}.
\eeq
Note that there are further anomalous dimensions, e.g. the one which belongs to the ghost fields or
the background anomalous dimension which appears due to the background field method
\cite{Manrique:2010mq,Christiansen:2012rx,Dona:2012am,Codello:2013fpa,Christiansen:2014raa,Becker:2014qya}.
The model has two fixed points \cite{Litim:2003vp,Fischer:2006fz,Litim:2006dx}.
When $d=4$, then there is a UV NGFP at $\lambda^*_{UV}=1/4$,
$g^*_{UV}=1/64$. The matrix $M$ is
\beq
M=\frac13\begin{pmatrix}
-2 & -176\\
-1 & -8\\
\end{pmatrix},
\eeq
with the eigenvalues $s_{UV1}=(-5+i\sqrt{167})/3$ and $s_{UV2}=(-5-i\sqrt{167})/3$,
so it is a UV attractive of IR repulsive focal point. The corresponding critical exponent is identified
as the negative reciprocal of the real part of the eigenvalues, thus $\nu=0.6$. Furthermore one has a GFP at
$\lambda^*_G=0$, $g^*_{UV}=0$ with the corresponding matrix
\beq\label{qegGM}
M=\frac13\begin{pmatrix}
-2 & \frac12 d(d+2)(d-3)\\
-1 & d-2\\
\end{pmatrix}
\eeq
in dimension d. The eigenvalues are $s_{G1}=-2$ and $s_{G2}=d-2$.
The negative reciprocal of the former eigenvalue gives the critical exponent $\nu=1/2$ of $\xi$.
The latter one guarantees that the GFP is a hyperbolic one when $d>2$.

\begin{figure}
\begin{center}
\epsfig{file=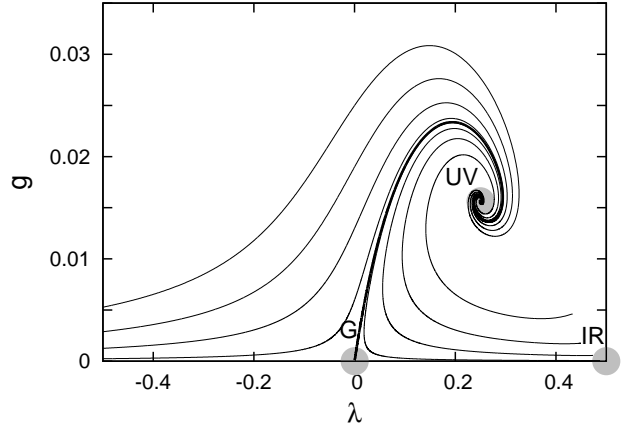,width=6cm,angle=-90}
\caption{\label{fig:qgfp} The phase structure of quantum Einstein gravity in $d=4$ is shown.
There is a UV attractive NGFP, a crossover fixed point, which is now the Gaussian one,
and an IR fixed point. The thick line represents the separatrix. The trajectories start from
the vicinity of the UV NGFP.}
\end{center}
\end{figure}

The phase structure is shown in \fig{fig:qgfp}.
There is a separatrix connecting the UV attractive NGFP and the GFP.
The trajectories which are left to the separatrix give negative
values for the cosmological constant and vanishing Newton coupling in the IR limit,
and constitute the strong-coupling or symmetric phase of QEG
\cite{Reuter:1996cp,Polonyi:2004ay}.
Other trajectories getting around the separatrix from the right give
large positive values of $\lambda$ and small Newton coupling when the RG flows approach
the IR regime. This phase is called the weak-coupling or broken symmetric phase.

In $d=4$ if one reparameterizes the couplings according to $\chi=1-2\lambda$,
$\omega=4g-(1-2\lambda)^2$ and introduces the new 'RG time'
$\partial_\tau=\omega \partial_t$ then one obtains
\bea
\partial_\tau\chi &=& -4\omega+2\chi\omega(8+21\chi)+24\omega^2+6\chi^2(3\chi(\chi+1)-1),\nn
\partial_\tau\omega &=& 8\omega^2(1-6\chi)-2\chi(42\chi^2+9\chi-4)-6\chi^3(\chi(6\chi+5)-2),
\eea
which have the UV NGFP at $\omega^*_{UV}=-3/16$ and $\chi^*_{UV}=1/2$, the GFP at
$\omega^*_G=-1$ and $\chi^*_G=1$ and the IR attractive fixed point at
$\omega^*_{IR}=0$ and $\chi^*_{IR}=0$ which corresponds to the point
$g^*_{IR}=0$ and $\lambda^*_{IR}=1/2$ in terms of the original dimensionless couplings.
Naturally the IR fixed point is IR attractive.
The existence of the IR fixed point also has been uncovered and discussed in
\cite{Donkin:2012ud,Litim:2012vz,Christiansen:2012rx}.
The RG equations of QEG coming from other truncation schemes can give distinct form
of propagators with different singularity conditions. In \cite{Nagy:2012rn} we showed
that in the IR evolution a scaling regime appears implying that there exists an IR
fixed point in those cases, too.

\begin{figure}
\begin{center} 
\epsfig{file=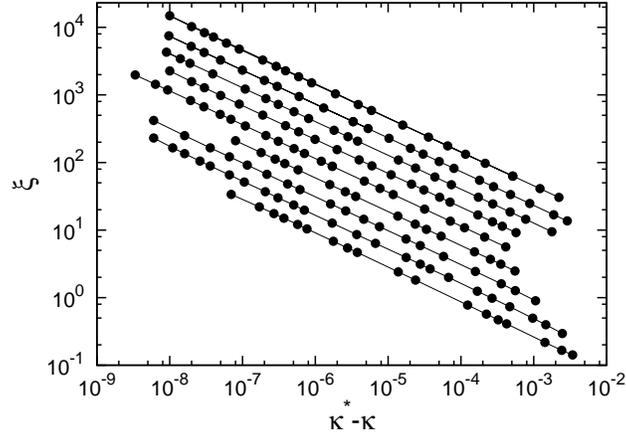,width=6cm,angle=-90}
\caption{\label{fig:qegcorr} The scaling of the correlation length, giving $\nu=1/2$
for the exponent. For simplicity we fixed the UV value of $g_\Lambda$.
} 
\end{center}
\end{figure}

As in the case of the previous models, one can get the value of the critical exponent $\nu$
from the IR scaling. Again, the scale $k$ where the evolution stops equals the
reciprocal of the correlation length $\xi$. We plotted the divergence of the correlation length
in \fig{fig:qegcorr} for dimensions $d=4\dots 11$.
We obtain that a second order phase transition can be identified in the IR region of
QED with the exponent $\nu=1/2$ independently on the dimension. One can obtain similar results
if one uses different regularization scheme or special extensions of the action containing
the functions of the Euclidean spacetime volume $V=\int d^dx\sqrt{\mbox{det}g_{\mu\nu}}$
\cite{Nagy:2012rn}. This result does not coincide to the UV value of $\nu=1/3$
\cite{Hamber:1991ba,Hamber:1999nu}, which is the reciprocal of the imaginary part
of the scaling exponent in the UV fixed point, i.e. $\nu=3/\sqrt{167}\approx 0.23$.
They are not equal but taking into account further couplings e.g. by including
higher order terms in the curvature scalar
\cite{Lauscher:2002sq,Codello:2006in,Fischer:2006at,Codello:2007bd,Benedetti:2010nr,
Groh:2011vn,Litim:2011cp} the result of $\nu$ can be improved.
Generally the exponent has some scheme and truncation dependence, and it is around
$\nu\approx 0.23-0.4$.

However the improvement of the action by extensions does not necessarily change the IR
value of $\nu=1/2$. As it was demonstrated through many models in the previous sections
the IR value of $\nu$ equals the one which can be calculated in the vicinity of the crossover
hyperbolic fixed point if it exists. Here the GFP plays this role, but we know
that the eigenvalues of the GFP equals the negative canonical dimension of the couplings.
The single negative eigenvalue of $M$ in \eqn{qegGM} is $s_1=-2$ independently
on the dimension, as is numerically checked and shown in \fig{fig:qegcorr}. If we introduce
new couplings into the QEG effective action in \eqn{eaqeg} as a Taylor expansion of the
dimensionless form of $\t f(\t R)$ \cite{Codello:2008vh} according to
\beq
\t f(\t R) = \sum_{n=1}^N \frac{\t g_n}{n!}\t R^n,
\eeq
then the canonical dimension of the coupling $\t g_n$ becomes $d_n=d-2n$. We can identify
the first two dimensionless couplings with the cosmological and the Newton couplings
as $\lambda\sim \t g_0/\t g_1$ and $g\sim 1/\t g_1$. Then the canonical dimension induced scaling
of the couplings around the GFP are $\lambda\sim k^{-2}$, $g\sim k^{d-2}$ and
$\t g_n\sim k^{2n-d}$ with $n>1$. In case of $d=4$ the single negative eigenvalue is
$s_1=-2$, which belongs to the cosmological coupling. It implies that the exponent
$\nu$ calculated at the GFP is always equals $1/2$, therefore we expect, that its IR
value is exactly $\nu=1/2$ if the total $\t f(\t R)$ is considered.

We note that there are extensions of QEG where the GFP does not exist \cite{Reuter:2002kd}.
Since the IR fixed point is strongly related to the crossover hyperbolic point then its
scaling behavior can significantly change. It was shown that a certain choice of the
extension may eliminate the GFP \cite{Nagy:2012rn}.  In this case the correlation length
$\xi$ does not diverge which implies that the order of the phase transition changes
to a first order one, in the IR region. Naturally it can also change the classical
limit there which may also affect the cosmological value problem \cite{Nagy:2012rn}.

\section{Summary}\label{sect:sum}

We gave a short introduction of asymptotic safety in the framework of the functional
renormalization group method. We introduced several models, where a nontrivial UV
attractive fixed point exists, which saves the model from nonphysical divergences
and gives asymptotic safety.

The functional renormalization group method is a powerful nonperturbative
tool to investigate the field theoretical models. It provides a partial differential
equation for the effective action which initial condition describes the high energy
UV physics of the model and the solution gives the low energy IR one. The RG technique
is used in several areas in quantum physics, however the results from its nonperturbative
nature comes forward quite rarely. One can say that the stable UV and the stable IR limit of the
theory cannot be described by perturbative calculations. The former one provides the
asymptotic safety, while the latter one gives the IR behavior of the broken phase
which can be related to the quantum-classical transition. In this articles we concentrated
on these limits of the models which were considered.

First we investigated the d-dimensional $O(N)$ model, which is asymptotically free.
We showed that there exist two phases and three fixed points in the model. Then we
analyzed such models which are asymptotically safe. We plotted the phase structure of
the 3d nonlinear $\sigma$, the 3d Gross-Neveu, the 2d sine-Gordon models and the 4d
quantum Einstein gravity. These models have two phases. The theory space shows several
similarities. There are three different fixed points. The trajectories start from the UV
attractive NGFP and tend towards the crossover hyperbolic fixed point. In the broken phase of the
models there exists a further IR fixed point which is IR attractive. \tab{tab:models}
summarizes the fixed points of the models, which were investigated.
\begin{table}
\begin{center}
\begin{tabular}{|c||c|c|c|}
\hline
{\bf model} & {\bf UV} & {\bf CO} & {\bf IR} \\
\hline
\hline
3d $O(N)$ model & Gaussian & Wilson-Fisher & IR \\
\hline
3d nonlinear $\sigma$ model & non-Gaussian & non-Gaussian & Gaussian \\
\hline
3d Gross-Neveu model & Gaussian & non-Gaussian & IR \\
\hline
2d sine-Gordon model & Gaussian and non-Gaussian & Kosterlitz-Thouless & Gaussian and IR \\
\hline
4d quantum Einstein gravity & non-Gaussian & Gaussian & IR \\
\hline
\end{tabular}
\end{center}
\caption{\label{tab:models} Summary of the models and their fixed points.}
\end{table}
The models are not in the same universality class, since they can be distinguished by the
critical exponents. Different fixed points may result in different exponents, therefore
the classification could be done for the UV, CO and the IR fixed points.

\section*{Acknowledgments}

This research was supported by the European Union and the State of Hungary, co-financed by the
European Social Fund in the framework of T\'AMOP 4.2.4.A/2-11-1-2012-0001 'National Excellence Program'.

\bibliography{nagy}

\end{document}